\documentclass[12pt]{article}
\usepackage{amssymb}
\usepackage{amsfonts}
\usepackage{amsmath}
\usepackage{harvard}
\usepackage{chicago}
\usepackage{theorem}
\usepackage{graphicx}

\newtheorem{theorem}{Theorem}

\newtheorem{corollary}{Corollary}

\newtheorem{lemma}{Lemma}

\newtheorem{remark}[theorem]{Remark}

\renewcommand{\cite}{\citeasnoun}
\renewcommand{\thepage}{}
\renewcommand{\thefootnote}{\fnsymbol{footnote}}
\allowdisplaybreaks
\renewcommand{\baselinestretch}{1.3}

\input{tcilatex}

\begin{document}

\title{Fixed-$k$ Inference for Conditional Extremal Quantiles\thanks{%
We thank Federico Bugni, Xiaohong Chen, Tim Christensen, Yanqin Fan, Yoonseok
Lee, Zhijie Xiao, Yichong Zhang, and participants at the seminar/conference
at Boston College, PSU, SMU, EC$^{2}$ Conference 2019, Econometric Society North American Summer Meeting 2019,
and Greater New York Metropolitan Area Econometrics Colloquium 2019, for
very helpful comments and advice. Wang gratefully acknowledges the financial
support by the Applyby-Mosher fund.}}
\author{Yuya Sasaki\thanks{
Associate professor of economics, Vanderbilt University. Email:
yuya.sasaki@vanderbilt.edu} \ and Yulong Wang\thanks{%
Assistant professor of economics, Syracuse University. Email:
ywang402@maxwell.syr.edu.}}
\date{First arXiv version: August 2019\\
This version: July 2020}
\maketitle

\begin{abstract}
\setlength{\baselineskip}{5.6mm} {\small We develop a new extreme value
theory for repeated cross-sectional and panel data to construct asymptotically
valid confidence intervals (CIs) for conditional extremal quantiles from a
fixed number $k$ of nearest-neighbor tail observations. As a by-product, we
also construct CIs for extremal quantiles of coefficients in linear random
coefficient models. For any fixed $k$, the CIs are uniformly valid without
parametric assumptions over a set of nonparametric data generating processes
associated with various tail indices. Simulation studies show that our CIs
exhibit superior small-sample coverage and length properties than
alternative nonparametric methods based on asymptotic normality. Applying
the proposed method to Natality Vital Statistics, we study factors of
extremely low birth weights. We find that signs of major effects are the
same as those found in preceding studies based on parametric models, but
with different magnitudes.  }

{\small { \ \ \newline
\textbf{Keywords: } conditional extremal quantile, confidence interval,
extreme value theory, fixed $k$, random coefficient} }
\end{abstract}

\setcounter{page}{1} \renewcommand{\thepage}{\arabic{page}} %
\renewcommand{\thefootnote}{\arabic{footnote}} 
\renewcommand{\baselinestretch}{1.2} \small \normalsize%
\newpage

\section{Introduction}

Tail risks and extreme events are important research topics in economics. In
many applications with multivariate analysis, features of interest are
conditional tail properties such as conditional extremal quantiles. This
article provides a new method to construct confidence intervals for
conditional extremal quantiles from a fixed number $k$ of nearest-neighbor tail observations.
Advantages of the proposed method are three-fold: first, it is robust
against flexible distributional assumptions unlike parametric methods; 
second, the procedure yields
asymptotically valid confidence intervals for any fixed tuning parameter $k$ unlike existing kernel methods that rely on sequences of moving tuning parameters for asymptotically valid inference; and third, our confidence intervals enjoy a uniform coverage property over a set of data generating processes involving a set of values of the tail index.
In the existing literature, methods of inference about conditional quantiles concern about middle quantiles, e.g., \cite{Qu2015} -- also see \cite{Qu2019} -- based on local quantile estimators of \cite{Fan1994} and \cite{Yu1998}.
We aim to complement this existing literature by proposing a method of inference about conditional extremal quantiles.

Compared with unconditional tail features, the conditional tail counterparts
are more difficult to study. This is because conditional tails depend on
both marginal distributions and their joint behavior. Although marginal
distributions can be generally assumed to be approximately Pareto near the
tails,\footnote{%
This statement follows from the Pickands-Balkema-de Haan Theorem (\cite%
{BalkemadeHaan1974} and \cite{Pickands1975}). See \cite{deHaan07} for an
overview.} joint distributions cannot be generally assumed to be
approximated by a fully parametric joint distribution and thus are harder to
study given very limited tail observations. To model a covariate-dependent
yet tractable tails, the seminal paper by \cite{Chernozhukov05} extends the
quantile regression (QR) estimator of \cite{Koenker78} to tails, and proposes a
method called the extremal quantile regression (EQR). \cite{Chernozhukov11}
further investigate the EQR to construct confidence intervals (CIs) based on
subsampling.

The EQR approach is based on the assumption that the conditional extremal quantile can be well approximated by a parametric location-scale shift model: 
\begin{equation}
Q_{Y|X=x}\left( \tau \right) \sim \mu \left( x\right) +\sigma \left(x\right)
(1-\tau )^{-\xi }  \label{EQR cond}
\end{equation}
for $\tau \rightarrow 1$, where $\mu \left( x\right) $ and $\sigma \left(x\right) $ are parametric functions that capture the location and scale, respectively. 
The element $(1-\tau )^{-\xi }$ can be treated as the quantile function of a standard Pareto distribution, that is, $\mathbb{P}\left( Y>y\right) \sim y^{-1/\xi }$ where $1/\xi $ is the Pareto exponent and $\xi $ is the tail index. 
This single parameter captures the tail shape in the way that a larger $\xi $ implies a heavier tail. 
The assumption of model (\ref{EQR cond}) simplifies the conditional tail distribution so that the covariate $X$ only affects the location and scale, but not the shape.\footnote{\cite{WangLi13} formally establish that the location-shift model
assumption is equivalent to assuming $\xi $ remains constant across $x$.}
This is satisfied if $X$ and $Y$ are jointly normal but violated by many
other joint distributions. Unlike mid-sample features, misspecification bias
could be substantial in studying tail ones.\footnote{With this said, we remark that the existing literature suggests a couple of ways in which one can rationalize a possibly misspecified quantile regression. \cite{Angrist06} show that the parametric linear quantile regression function minimizes a weighted distance to the true nonparametric quantile regression function. \cite{Kato17} show that the linear quantile regression parameter is a weighted average of the slopes of the true
nonparametric quantile regression function.}
In this paper, we consider a wider class of flexible joint distribution models using a repeated cross-sectional or panel data structure.

There are a number of reasons for which we want to study conditional tail features, such as conditional extremal quantiles, under flexible joint distribution models. First, conditional value-at-risk (VaR) is a risk measure commonly used in financial management, insurance, and actuarial science. 
Estimation and inference are studied by \cite{Chernozhukov01} and \cite{Engle04}, among others. 
\cite{Adrian16} propose a new measure for systemic risk, $\Delta $-CoVar, defined as the difference between two conditional VaRs. 
The tail shape governs the third-and higher-order moments of the portfolio return, which typically depend on other economic factors, e.g., business cycles. 
As this is excluded by the location-scale model (\ref{EQR cond}), it is preferred to accommodate a larger class of joint distributions. 
Second, \cite{Kelly14} find that extreme event risk affects asset pricing in the U.S. stock market. 
The shape parameter measures tail risk and varies with other stock characteristics such as stock size.
Third, macroeconomists are interested in analyzing lower tails of the conditional distributions of GDP growth rate given financial conditions in the recent growth-at-risk literature -- see \cite{Adrian19} for example.
Fourth, top wealth inequality is an active research question in macro-finance literature (see, for example, \cite{Piketty03}, \cite{Gabaix16}, and \cite{Jones18}). 
The tail of the wealth distribution is well documented to follow Pareto, and the exponent is in general a function of fundamentals in general equilibrium models. 
For example, \cite{Beare17} derive a formula for the Pareto exponent and comparative statics results, and \cite{Toda19} applies that formula in a general equilibrium context.
Finally, investigating factors of infants' birth weights, such as mother's demographic characteristics and maternal behaviors, is an important question
in health economics (e.g., \cite{Abrevaya01} and \cite{Koenker01}). 
The lower tails of the conditional distribution are especially of interest for
their critical health consequences -- see \cite{Chernozhukov11}. 
Other economic issues about conditional tail features can be found in the
comprehensive review by \cite{Chernozhukov17}.

The existing literature suggests alternative approaches besides those based
on the parametric location-scale specification (\ref{EQR cond}). To our best
knowledge, they all focus on estimation, as opposed to inference, and can be
roughly categorized into two classes. The first class maintains some
parametric form but relaxes the location-shift model to allow for some
nonlinearity. \cite{WangTsai09} assume that $\xi (x)$ equals to $%
\exp(x^{\intercal }\theta_{0}) $ for some unknown parameter $\theta _{0}$. 
\cite{WangLi13} assume that the Box-Cox transformed $Y$ has linear
conditional quantiles in $X$. The second class is fully nonparametric and
constructs some local smooth estimators, including, for example, \cite%
{Beirlant04}, \cite{Gardes10}, \cite{Gardes12}, \cite{Daouia13}, and \cite%
{Martins-Filho18}. 

In this article, we focus on statistical inference rather than estimation,
and provide confidence intervals (CIs) of a conditional extremal quantile
that have preferred coverage and length properties. Our proposed method
applies to both repeated cross-sectional data and panel data. The main idea is
very intuitive. Consider the case of using panel data of $(Y,X)$ to fix
ideas, and suppose that one is interested in the conditional extremal
quantile of $Y $ given $X=x_{0}$, denoted by $Q_{Y|X=x_{0}}(\tau)$. If, for
every individual, there exists some time period in which $X$ takes the value $x_{0}$,
then we can simply collect the associated $Y$'s and form a cross-sectional
sample from $F_{Y|X=x_{0}}$. Since this is infeasible especially when $X$ is
continuous, we instead collect from each individual's time series the
induced $Y$ associated with $X$ that is the nearest neighbor (NN) of $x_{0}$%
. These induced $Y$'s are now \textit{approximately} stemming from $%
F_{Y|X=x_{0}}$, and the large (respectively, small) order statistics from
them can be used for inference about the upper (respectively, lower)
conditional extremal quantile $Q_{Y|X=x_{0}}(\tau)$. For multi-dimensional
covariates, this is done by defining the NN measured by a certain choice of
metric, such as the one induced by the Euclidean norm. If a linear
regression model is appropriate, then the NN can also be defined using the
linear index.

The above approximation approach is formalized by establishing a new extreme
value (EV) theory. The theory is based on the large-$n$ and large-$T$
asymptotics, where $n$ and $T$ denote the sample sizes in cross-sectional
and time-series dimensions, respectively. A large $T$ guarantees that the NN
is close enough to the query point $x_{0}$, while a large $n$ provides
enough observations from a more accurate tail sample. Given the new EV
theory, we apply it to construct new confidence intervals for the conditional extremal
quantiles.

Our proposed approach only requires some smoothness condition on the joint
distribution and hence enjoys more robustness against functional form
specification than existing methods. A natural question is how much
efficiency we lose by using only one out of $T$ observations in each time
series. It turns out that if the tail shape depends on the covariate highly
nonlinearly,\footnote{%
See Section \ref{sec MC} for concrete numerical settings which this
qualitative phrase stands for.} then our proposed NN method dominates
existing methods in both coverage and length when $T$ is only moderately
large, say 50. When $T$ is very large, say 500, the new CIs also deliver
comparable lengths to the kernel regression method with the optimal
bandwidth -- see the Monte Carlo results ahead in Section \ref{sec MC} for
more details.

As a by-product of our main result, we also develop CIs for extremal quantiles of the coefficients in a random coefficient regression model. 
In particular, suppose that $Y_{it}$ and $X_{it}$ are generated from the model $Y_{it}=\alpha _{i}+X_{it}^{\intercal }\beta _{i}+u_{it}$, where $(\alpha_{i},\beta _{i}^{\intercal })^{\intercal}$ is a random vector drawn from some unknown distribution. We first construct the least squares estimators of $\alpha_{i} $ and $\beta _{i}$ using the $i$-th time series for all $i$ and collect the largest (smallest) order statistics from these estimates. 
We then show that the estimation error is negligible under the large $n$ and large $T$
framework, and hence the largest (smallest) order statistics among these estimates again satisfy the desired EV theory, which further supports the application of the fixed-$k$ CIs for extremal quantiles of $\alpha _{i}$ and $\beta _{i}$. 
This complements the existing literature focusing on the mid-sample properties of heterogeneous effects (e.g., \cite{Hsiao04} and  \cite{Wooldridge05}).

Applying the proposed methods, we study the tail risk of extremely low birth weight conditional on mothers' behavioral and demographics characteristics. 
We find that signs of major effects are the same as those found in preceding studies based on parametric models. 
On the other hand, we find that some effects exhibit different magnitudes from those reported in the previous studies based on parametric models.

The rest of the paper is organized as follows. 
Section \ref{sec main} presents the main results of this paper.
Section \ref{sec MC} presents Monte Carlo simulation studies. 
Section \ref{sec:application} presents an empirical application. 
Section \ref{sec conclusion} concludes the paper. 
All mathematical proofs and additional details are found in the appendix.

\textbf{Notation} Let $\overset{p}{\rightarrow }$ denote convergence in
probability and $\overset{d}{\rightarrow }$ denote convergence in
distribution as $n,T\rightarrow \infty $. Let $\mathbf{1}[A]$ denote the
indicator function of a generic event $A$. Let $\left\vert \left\vert
B\right\vert \right\vert $ denote the Euclidean norm of a vector or matrix $%
B $, and let $C$ denote a generic constant whose value may change across
lines. Let $B_{\delta }(x)$ denote a generic open ball centered at $x$ with
radius $\delta $. When $X$ denotes a column vector and $c$ a scalar, the
notation $X-c$ is understood as the vector $X-(c,\ldots ,c)^{\intercal }$.

\section{Main result\label{sec main}}

We present the main result of this paper in this section. Let $X$ denote a $%
\dim (X)\times 1$ vector of continuous random variables with uniformly
positive joint PDF.\footnote{%
While we focus on continuous random variables in our presentation, the
method can also accommodate discrete random variables. Suppose that the
covariate vector is written as $(X^{\prime },W^{\prime })^{\prime }$ where
the subvector $X$ consists of continuous random variables and the subvector $%
W$ consists of discrete random variables. Suppose that one is interested in
the conditional extremal quantiles given $(X^{\prime },W^{\prime })^{\prime
}= (x_0^{\prime },w_0^{\prime })^{\prime }$. We can then extract the
subsample with $W=w_0$, and then apply our proposed method for the subsample.%
} The main object of interest is the conditional extremal quantile $%
Q_{Y|X=x_{0}}\left( \tau \right) $ of $Y$ given $X = x_{0}$ for a
pre-specified $x_{0}\in \mathbb{R} ^{\dim \left( X\right) }$ and $\tau
\rightarrow 1$. For ease of exposition, we consider a balanced\footnote{%
This is only for notational ease. The new approach is valid as long as $T$
is large for all $i$.} repeated cross-sectional or panel data set $\{Y_{it},X_{it}\}_{i=1:n,t=1:T}$ that
is i.i.d. across $i$ and strictly stationary and weakly dependent across $t$%
. Section \ref{sec general} presents an informal overview of our proposed
method, and Section \ref{sec condition and result} gives a formal
theoretical justification.
Finally, Section \ref{sec ext} presents an extension of the main theoretical results to linear random coefficient models.

\subsection{Overview\label{sec general}}

Our method consists of the following three steps. In the first step, we
make use of the repeated cross-sectional or panel data structure by selecting a subsample induced by the distances of the covariates $\{X_{it}\}$ to the query point $x_{0}$. 
The subsample will be a $k\times 1$ random vector denoted as $\mathbf{Y}$. In
the second step, by appealing to the extreme value theory, we show that
after some normalization, $\mathbf{Y}$ converges in distribution to a
well-defined limiting random variable, $\mathbf{V}$, whose distribution $f_{%
\mathbf{V}}$ is parametric and uniquely determined by the tail features of $%
F_{Y|X=x_{0}}$. In particular, $f_{\mathbf{V}}$ will be uniquely
characterized by a scalar parameter $\xi $ that fully captures the tail
heaviness of $F_{Y|X=x_{0}}$. Note that $\xi $ depends on the query point $%
x_{0}$, which will be suppressed in our notations for simplicity when there
is no confusion. Since $Q_{Y|X=x_{0}}\left( \tau \right) $ can also be
uniquely expressed as a function of $\xi $ after suitably normalizing $\tau $%
, the asymptotic problem becomes conceptually straightforward: constructing
inference for a function of $\xi (x_{0})$ given a random draw $\mathbf{V}$.
This type of problems is studied by \cite{EMW15}, who provide a generic
argument to construct optimal inference when there exists a nuisance
parameter under a null hypothesis. In the third and final step, we tailor
their arguments to inference about $Q_{Y|X=x_{0}}\left( \tau \right) $ with $%
\xi $ being the nuisance parameter.

The next three subsubsections introduce details of these three steps in order. Section \ref{sec condition and result} then follows up
by presenting regularity conditions and the main theoretical result of the
paper that guarantees that our confidence interval constructed in the
three-step procedure controls coverage asymptotically and uniformly over a set of data generating processes.

\subsubsection{Step 1: subsample selection based on NN}

First, we select our subsample $\mathbf{Y}$ as follows.

\begin{itemize}
\item Collect, for each $i$, the induced $Y$ associated with the NN of $%
\{X_{it}\}_{t=1}^{T}$ to $x_{0}$, where the NN is measured by the Euclidean
distance $\left\vert \left\vert X_{it}-x_{0}\right\vert \right\vert$. Denote
them by $\{Y_{i,[x_{0}]}\}_{i=1}^{n}$.\footnote{%
Details of this step with more notations are as follows. For each $i \in
\{1,...,n\}$ and $t \in \{1,...,T\}$, compute $d_{it} = \|X_{it}-x_0\|$
where $\|\cdot\|$ denotes the Euclidean distance.  Then, for each $i \in
\{1,...,n\}$, let $t_i^\ast$ denote the argument $t$ that minimizes $d_{it}$%
. We denote $Y_{i,[x_0]} = Y_{it_i^\ast}$.}

\item Take the largest $k$ order statistics from $\{Y_{i,[x_0]}\}_{i=1}^{n}$
and denote the vector of them by 
\begin{equation}
\mathbf{Y}=\mathbf{(}Y_{(1),[x_{0}]},Y_{(2),[x_{0}]},...,Y_{(k),[x_{0}]})^{%
\intercal },  \label{Y_k}
\end{equation}
where $Y_{(1),[x_{0}]}\geq Y_{(2),[x_{0}]}\geq \ldots \geq Y_{(n),[x_{0}]}$
are the order statistics of $\{Y_{i,[x_{0}]}\}_{i=1}^{n}$.
\end{itemize}

The key idea for such a selection is heuristically illustrated by the
following derivation - a formal argument is presented as a proof of Theorem \ref{thm
main} in Appendix \ref{sec:proof}. For each $i$, denote the NN among $%
\{X_{it}\}_{t=1}^{T} $ to $x_{0}$ as $X_{i,(x_{0})}$. Then for any $y\in 
\mathbb{R} $, 
\begin{align*}
&\mathbb{P}\left( Y_{i,[x_{0}]}\leq y\right) \\
=&\mathbb{E}_{X_{i,\left( x_{0}\right) }}\left[ \mathbb{P}\left(
Y_{i,[x_{0}]}\leq y|X_{i,\left( x_{0}\right) }\right) \right] \\
=&\mathbb{E}_{X_{i,\left( x_{0}\right) }}\left[ F_{Y|X=X_{i,\left(
x_{0}\right) }}\left( y\right) \right] &&\text{ (by strict stationarity)} \\
=&F_{Y|X=x_{0}}\left( y\right) +\mathbb{E}_{X_{i,\left( x_{0}\right) }}%
\left[ \left. \frac{\partial F_{Y|X=x}\left( y\right) }{\partial
x^{\intercal }}\right\vert _{x=\dot{x}_{i}}(X_{i,\left( x_{0}\right) }-x_{0})%
\right] &&\text{ (by mean value expansion)} \\
\rightarrow &F_{Y|X=x_{0}}\left( y\right) &&\text{ as }T\rightarrow \infty ,
\end{align*}%
where $\dot{x}_{i}$ lies between $X_{i,\left( x_{0}\right)}$ and $x_{0}$.
The first equality is by the definition of conditional expectation. The
second one follows under the strict stationarity. The third equality is
valid if the conditional CDF is smooth. The last convergence holds if the NN
converges to its query point $x_{0}$ and if the CDF is smooth with
bounded derivatives.

The above derivation states that the collection of the induced order
statistics $Y$ associated with the NN of $x_{0}$ can be treated as
approximately stemming from the true conditional CDF $F_{Y|X=x_{0}}$
asymptotically. Thus the largest (cross-sectional) order statistics $\mathbf{%
Y}$ can be treated as draws from the tail of $F_{Y|X=x_{0}}$.

\subsubsection{Step 2: asymptotic distribution of the subsample}

To proceed with the second step, we need some regularity conditions about $%
F_{Y|X=x_{0}}$. For readability, we introduce only one of the conditions
here with the remaining of them discussed in Section \ref{sec condition and
result}. Specifically, we assume that $F_{Y|X=x_{0}}$ is within the domain
of attraction (DOA) of the extreme value distribution (denoted by $%
F_{Y|X=x_{0}}\in \mathcal{D}(G_{\xi })$), in the sense that there exist
sequences of constants $a_{n}$ and $b_{n}$ such that for every $v$, 
\begin{equation*}
\lim_{n \rightarrow \infty}F_{Y|X=x_{0}}(a_{n}v+b_{n})= G_{\xi }(v),
\end{equation*}
where 
\begin{equation}
G_{\xi }(v)=\left\{ 
\begin{array}{ll}
\exp (-(1+\xi v)^{-1/\xi })\text{, } & 1+\xi v>0\text{, for }\xi \neq 0 \\ 
\exp (-e^{-v})\text{, } & v\in \mathbb{R} \text{, }\xi =0.%
\end{array}
\right.  \label{def_G}
\end{equation}
This DOA condition is extensively studied in the statistics literature and
is satisfied by many commonly used distributions, including, for example,
Pareto, Student-t, F, Gaussian, and even uniform distributions. See Chapter
1 in \cite{deHaan07} for a complete review.

Under the DOA assumption and the cross-sectional i.i.d. assumption, we show
that for any fixed $k$, 
\begin{equation}
\frac{\mathbf{Y-}b_{n}}{a_{n}}\overset{d}{\rightarrow }\mathbf{V}=\left( 
\begin{array}{c}
V_{1} \\ 
\vdots \\ 
V_{k}%
\end{array}%
\right) \text{, as }n,T\rightarrow \infty ,  \label{evt_k}
\end{equation}%
where the joint probability density function (PDF)~of $\mathbf{V}$ is given
by 
\begin{equation}
f_{\mathbf{V}}(v_{1},\ldots ,v_{k};\xi )=G_{\xi
}(v_{k})\prod_{i=1}^{k}g_{\xi }(v_{i})/G_{\xi }(v_{i})  \label{evt_pdf}
\end{equation}%
for $v_{k}\leq v_{k-1}\leq \ldots \leq v_{1}$ with $g_{\xi }(v)=\partial
G_{\xi }(v)/\partial v$, and zero otherwise.

Note that the constants $a_{n}$ and $b_{n}$ depend on $\xi $, and their
estimates thus tend to exhibit large magnitudes of sensitivity to tail
observations. For example, $a_{n}$ is $n^{\xi }$ if $F_{Y}$ is standard
Pareto. Since a small estimation error in $\xi $ is amplified by the $n$%
-power, inference relying on a good estimate of $\xi $ and the scale usually
requires a large $k$ and a even larger sample size $n$. Besides, the $G_{\xi
}(v_{k})$ term in (\ref{evt_pdf}) suggests that the largest $k$ order
statistics are not asymptotically independent, given any fixed $k$.
\footnote{We also derived the estimation and inference method based on the increasing-$k$ asymptotics in a previous version of this article. Their performance is dominated by the fixed-$k$ approach, especially in case with only moderate sample sizes. Therefore, we present the fixed-$k$ result exclusively in this version for illustrational simplicity.}

\subsubsection{Step 3: construction of the asymptotic inference}

We aim for a $(1-\alpha)$ CI for the conditional extremal quantile $%
Q_{Y|X=x_{0}}(\tau )$ for $\tau $ close to 1. Specifically, we rewrite $\tau$
as $1-h/n$ for some $h>0$ following \cite{Chernozhukov05} and \cite%
{Chernozhukov11}. This setup means that the extremal quantile is of the same
order of the sample maximum from $n$ random draws from the true conditional
CDF $F_{Y|X=x_{0}}$.

Our objective is to construct a confidence set $S(\mathbf{Y})\subset \mathbb{%
R} $ such that $\mathbb{P}(Q_{Y|X=x_{0}}(\tau )\in S(\mathbf{Y}))\geq
1-\alpha + o(1)$, as $n\rightarrow \infty $ and $T\rightarrow \infty $.
Under the DOA assumption, calculations show that 
\begin{equation*}
\frac{Q_{Y|X=x_{0}}(1-h/n)-b_{n}}{a_{n}}\rightarrow q(\xi ,h)\equiv \left\{ 
\begin{array}{ll}
\frac{h^{-\xi }-1}{\xi } & \text{if }\xi \neq 0 \\ 
-\log (h) & \text{if }\xi =0.%
\end{array}%
\right.
\end{equation*}%
Note that $q(\xi ,h)$ is the $\exp (h)$ quantile of $V_{1}$. Since it is
shared by both $\mathbf{Y}$ and $Q_{Y|X=x_{0}}(1-h/n)$, we can impose
location and scale equivariance on the CI to cancel them out. Specifically,
we impose that for any constants $a>0$ and $b$, $S(a\mathbf{Y}+b)=aS(\mathbf{%
Y})+b$, where $aS(\mathbf{Y})+b=\{y:(y-b)/a\in S(\mathbf{Y})\}$. Under this
equivariance constraint, we can write 
\begin{eqnarray*}
&&\mathbb{P}(Q_{Y|X=x_{0}}(1-h/n)\left. \in \right. S(\mathbf{Y})) \\
&=&\mathbb{P}\left( \frac{Q_{Y|X=x_{0}}(1-h/n)-Y_{(k),[x_{0}]}}{%
Y_{(1),[x_{0}]}-Y_{(k),[x_{0}]}}\in S\left( \frac{\mathbf{Y}-Y_{(k),[x_{0}]}%
}{Y_{(1),[x_{0}]}-Y_{(k),[x_{0}]}}\right) \right) \\
&\rightarrow &\mathbb{P}_{\xi }\left( V^{q}\in S(\mathbf{V}^{\ast })\right),
\end{eqnarray*}
where we introduce the self-normalized statistics 
\begin{eqnarray*}
V^{q} &=&\frac{q(\xi ,h)-V_{k}}{V_{1}-V_{k}} \\
\mathbf{V}^{\ast } &\mathbf{=}&\left( \frac{V_{1}-V_{k}}{V_{1}-V_{k}},\frac{%
V_{2}-V_{k}}{V_{1}-V_{k}},...,\frac{V_{k}-V_{k}}{V_{1}-V_{k}}\right) ,
\end{eqnarray*}
and highlight with the subscript $\xi $ that the densities of $V^{q}$ and $%
\mathbf{V}^{\ast }$ now depend solely on $\xi $. These can be computed by
using (\ref{evt_k}), (\ref{evt_pdf}), and a change of variables.

Since $\xi $ is unknown, we impose the size constraint uniformly for all the
values of $\xi $ that are empirically relevant. In this sense the fixed-$k$
approach is more robust against misspecification, especially when the sample
size is not large enough to support a precise estimation of $\xi$. Let $\Xi
\subset\mathbb{R}$ be the set of tail indices for which we impose the
asymptotically correct coverage.\footnote{%
We use\ $\Xi =[-1/2,1/2]$ for inference about conditional extremal quantiles
in later applications, which covers all the distributions with finite
variance. This range can be easily extended.} The asymptotic problem then is
to construct a location and scale equivariant $S$ that satisfies 
\begin{equation}
\mathbb{P}_{\xi }\left( V^{q}\in S(\mathbf{V}^{\ast })\right) \geq 1-\alpha 
\text{ for all }\xi \in \Xi ,  \label{asy_cov}
\end{equation}
since any $S$ that satisfies (\ref{asy_cov}) also satisfies $\lim
\inf_{n\rightarrow \infty ,T\rightarrow \infty }\mathbb{P}%
(Q_{Y|X=x_{0}}(1-h/n)\in S(\mathbf{Y}))\left. \geq \right. 1\left.
-\right.\alpha $ by the continuous mapping theorem. Among all the solutions
to this problem, we choose the optimal one that minimizes the weighted
average expected length criterion 
\begin{equation}
\int \mathbb{E}_{\xi }[\text{lgth}(S(\mathbf{V}))]dW(\xi )\text{,}
\label{asy_length}
\end{equation}
where $W$ is a positive measure with support on $\Xi $,\footnote{%
We use the uniform weight in later sections.} and $\text{lgth}(A)=\int 
\mathbf{1}[y\in A]dy$ for any Borel set $A\subset \mathbb{R}$. The
equivariance of $S$ further implies $\mathbb{E}_{\xi }[\text{lgth}(S(\mathbf{%
V}))]\left. =\right. \mathbb{E}_{\xi }[(V_{1}-V_{k})\text{lgth}(S(\mathbf{V}%
^{\ast }))]$. Thus the program of minimizing (\ref{asy_length}) subject to (%
\ref{asy_cov}) among all equivariant sets $S$ asymptotically becomes 
\begin{equation}
\begin{tabular}{l}
$\min_{S(\cdot )}\int_{\Xi }\mathbb{E}_{\xi }[(V_{1}-V_{k})\text{lgth}(S(%
\mathbf{V}^{\ast }))]dW(\xi )$ \\ 
\multicolumn{1}{r}{$\text{s.t. }\mathbb{P}_{\xi }(V^{q}\in S(\mathbf{V}%
^{\ast }))\geq 1-\alpha \text{ for all }\xi \in \Xi ,$}%
\end{tabular}
\label{program}
\end{equation}%
where we abuse the notation of $\mathbb{E}_{\xi }$ and $\mathbb{P}_{\xi }$
to emphasize that the distributions of $V^{q}$ and $\mathbf{V}^{\ast }$
depend on $\xi $ (and further on $x_{0}$). Note that any solution to (\ref%
{program}) also provides the form of $S$, that is, $S(\mathbf{V}%
)=(V_{1}-V_{k})S(\mathbf{V}^{\ast })+V_{k}$. Once $S(\cdot )$ is determined,
therefore, the confidence interval can be constructed in practice by
plugging in 
\begin{equation*}
(Y_{(1),[x_{0}]}-Y_{(k),[x_{0}]})S\left( \frac{\mathbf{Y}-Y_{(k),[x_{0}]}%
}{Y_{(1),[x_{0}]}-Y_{(k),[x_{0}]}}\right) +Y_{(k),[x_{0}]}.
\end{equation*}

In solving (\ref{program}), we write the problem in the following Lagrangian
form: 
\begin{equation*}
\min_{S(\cdot )}\int_{\Xi }\mathbb{E}_{\xi }[(V_{1}-V_{k})\text{lgth}(S(%
\mathbf{V}^{\ast }))]dW(\xi )+\int_{\Xi }\mathbb{P}_{\xi }\left( V^{q}\in S(%
\mathbf{V}^{\ast })\right) d\Lambda (\xi ),
\end{equation*}%
where the non-negative measure $\Lambda $ denotes the Lagrangian weights
that guarantee the asymptotic coverage constraint. By defining $\kappa (\mathbf{V%
}^{\ast };\xi )=\mathbb{E}_{\xi }[V_{1}-V_{k}|\mathbf{V}^{\ast }]$ and
writing the expectations above as integrals over the densities $f_{\mathbf{V}%
^{\ast }}$ and $f_{V^{q},\mathbf{V}^{\ast }}$ of $\mathbf{V}^{\ast }$ and $%
(V^{q},\mathbf{V}^{\ast })$, respectively, the solution of the above problem is given by 
\begin{equation}
S(\mathbf{v}^{\ast })=\left\{ y:\int_{\Xi }\kappa (\mathbf{v}^{\ast };\xi
)f_{\mathbf{V}^{\ast }}(\mathbf{v}^{\ast };\xi )dW(\xi )<\int_{\Xi }f_{V^{q},%
\mathbf{V}^{\ast }}(y,\mathbf{v}^{\ast };\xi )d\Lambda (\xi )\right\} .
\label{S_Lambda}
\end{equation}%
The integrals can be numerically computed by Gaussian quadrature. To find
suitable Lagrangian weights $\Lambda $, we appeal to the generic algorithm
in \cite{EMW15}, who provide a numerical method to construct $\Lambda $. We
tailor their arguments to our conditional extreme tail inference problem and
provide the corresponding MATLAB program on the author's website. The
computation cost is only several seconds using a modern PC. Note that $%
\Lambda $ only needs to constructed once by the author but not the empirical
users. See Section \ref{sec computation} for more details.

Other tail-related quantities, such as the conditional tail expectations,
are also covered by our proposed method as long as they can be expressed as
functions of the conditional tail index. We discuss such an extension in
Section \ref{sec ext}. In the following subsection, we formally introduce all the
regularity conditions and formalize the uniform coverage property of the
confidence interval (\ref{S_Lambda}).

\subsection{Conditions and main theoretical results\label{sec condition and
result}}

Our asymptotic theory requires the following four conditions.

\begin{description}
\item[Condition 1.1] $(Y_{i1},X_{i1}^{\intercal })^{\intercal
},\ldots,(Y_{iT},X_{iT}^{\intercal })^{\intercal }$ are i.i.d.\ across $i$. $%
(Y_{it},X_{it}^{\intercal })^{\intercal }$ for each $t=1,\ldots ,T$ is
strictly stationary and $\beta $-mixing with the mixing coefficient
satisfying $\beta \left( t\right) =O(t^{-2-\varepsilon })$ for some $%
\varepsilon >0$. In addition, $f_{X}(x)$ is uniformly continuously
differentiable and bounded away from $0$ in an open ball centered at $x_{0}$.
\end{description}

Condition 1.1 requires the data to be independent across $i$ and weakly
dependent across $t$, which is plausibly satisfied by the Natality Vital
Statistics that we use for our empirical application. In addition, this condition also
requires the density of $X$ to be positive in an open neighborhood around
the query point $x_{0}$. This condition is sufficient to establish that the
NN converges to the query point $x_{0}$ almost surely at some power rate. To
the best of our knowledge, this is the first result about the (almost sure
and L$^{2}$) convergence rate of the NN under weak dependence, whose proof
is non-trivial. We formalize this result as Lemma \ref{lemma NN} in Appendix
A.1, which might be of independent research interest.\footnote{%
The $\beta $-mixing condition allows the application of Berbee's lemma (\cite%
{Berbee87}) in establishing Lemma \ref{lemma NN}. This is assumed to avoid
technical complexity and can be relaxed to other forms of weak dependence.}
Note that we intentionally choose only one NN to allow for weak dependence
across $t$. If data are independent across both $i$ and $t$, then more than
one NNs can be chosen to enlarge the effective sample. We leave this for future research.

\begin{description}
\item[Condition 1.2] $F_{Y|X=x_{0}}\in \mathcal{D}\left( G_{\xi \left(
x_{0}\right) }\right) $ with $\xi (x_{0})\in \Xi \,$, a compact subset of $%
\mathbb{R}
$.
\end{description}

This condition requires that the underlying conditional distribution is in
the domain of attraction of the generalized EV distribution. This is a mild
condition as it is satisfied by many commonly used joint distributions. In
particular, it generalizes the conditional location-scale shift model (\ref%
{EQR cond}) by allowing $\mu (x),\sigma (x),$ and $\xi (x)$ to be all
unknown (but smooth) functions of $x$. The case of negative $\xi (x_{0})$ is
included only for comprehensiveness, since the $Y$ in most applications involving
tail features has an unbounded support that entails a non-negative $\xi(x_{0})$. To illustrate the mildness of this
condition, we discuss the following three examples. Our Condition 1.2 is
satisfied in all three of them, but the location-scale model assumption (\ref%
{EQR cond}) is not.

\begin{description}
\item[Example 1 (Joint Normal)] Suppose that $(Y,X)$ is jointly normal with
zero means, unit variances, and correlation $\rho $. Then the conditional
distribution of $Y$ given $X=x$ is normal with mean $\rho x$, and variance $%
1-\rho ^{2}$. The conditional tail index is $\xi (x)=0$ for all $x\in 
\mathbb{R}$. The conditional quantile is $Q_{Y|X=x}\left( \tau \right) =\rho
x+\sqrt{1-\rho ^{2}}\Phi ^{-1}\left( \tau \right) $, where $\Phi ^{-1}\left(
\cdot \right) $ is the quantile function of the standard normal
distribution. Thus, the location-scale model assumption (\ref{EQR cond}) is
satisfied.

\item[Example 2 (Joint Student-t)] Suppose that $(Y,X)$ is jointly Student-t
distributed with d.f.$\ v$, zero means, unit variances, and correlation $%
\rho \neq 0$. Then the conditional distribution of $Y$ given $X=x$ is
Student-t distributed with d.f.\ $v+1$, mean $\rho x$, and variance $(1-\rho
^{2})(v+x^{2})/(v+1)$. The conditional tail index is $\xi (x)=1/(v+1)$ for
all $x\in \mathbb{R}$.\footnote{%
See \cite{Ding16} for the exact expression for the PDF.} The conditional
quantile is $Q_{Y|X=x}\left( \tau \right) =\rho x+\sqrt{(1-\rho
^{2})(v+x^{2})/(v+1)}Q_{t(v)}(\tau )$, where $Q_{t(v)}(\cdot )$ is the
quantile function of the standard Student-t distribution with d.f. $v$. This
specification satisfies the location-scale shift model (\ref{EQR cond}) but
the scale function is highly nonlinear in $x$.

\item[Example 3 (Conditional Pareto)] Suppose that $X$ is half-normal with
positive support and $Y$ given $X=x$ is the Pareto distribution such that $%
\mathbb{P}(Y\leq y|X=x)=1-(y+1)^{-1/x}$ for $y\geq 0$ and any $x>0$. Then
the conditional tail index is $\xi (x)=x$ and the conditional quantile is $%
Q_{Y|X=x}\left( \tau \right) =-1+(1-\tau )^{-x}$, which violates the
location-scale shift model (\ref{EQR cond}).
\end{description}

Let $y_{0}$ denote the end-point of the conditional CDF, that is, $%
y_{0}=Q_{Y|X=x_{0}}\left( 1\right) \leq \infty $. The next condition is a
high level regularity assumption on the smoothness of the conditional tail.

\begin{description}
\item[Condition 1.3] $f_{Y|X=x}(y)$ is uniformly bounded and continuously
differentiable in $x$ and $y$. In addition, for any fixed $y\left. >\right.
0 $ with $u_{n}\left. =\right. a_{n}y\left. +\right. b_{n}\left. \rightarrow
\right. y_{0}$, and any open ball $B_{\eta _{T}}\left( x_{0}\right) $
centered at $x_{0}$ with radius $\eta _{T}\left. \equiv \right. O(T^{-\eta
}) $ for some $\eta \left. >\right. 0$, $\lim_{u_{n}\rightarrow y_{0}}\left.
\sup_{x\in B_{\eta _{T}}\left( x_{0}\right) }\right. \left. T^{-\eta
}\left\vert \left\vert \frac{\partial F_{Y|X=x}\left( u_{n}\right) /\partial
x}{1-F_{Y|X=x_{0}}\left( u_{n}\right) }\right\vert \right\vert \right.
\left. =\right. 0$ and $\lim_{u_{n}\rightarrow y_{0}}\left. \sup_{x\in
B_{\eta _{T}}\left( x_{0}\right) }\right. \left. T^{-\eta }\left\vert
\left\vert \frac{\partial f_{Y|X=x}\left( u_{n}\right) /\partial x}{%
f_{Y|X=x_{0}}\left( u_{n}\right) }\right\vert \right\vert \right. \left.
=\right. 0$ as $n\left. \rightarrow \right. \infty $ and $T\left.
\rightarrow \right. \infty .$
\end{description}

Condition 1.3 requires that the derivatives of the conditional CDF and PDF
are smooth and decay quickly. This is a mild condition again, which is
satisfied by the above examples by straightforward calculation. For
readability, we provide low-level primitive assumptions as sufficient
conditions for Condition 1.3 and discuss them in Appendix \ref{sec:primitive_conditions}.

\begin{description}
\item[Condition 1.4] $n\rightarrow \infty $, $T\rightarrow \infty $, and $%
T/n\rightarrow \lambda $ for some $\lambda \in (0,\infty )$.
\end{description}

Condition 1.4 requires both $n$ and $T$ to be large. A large $n$ guarantees
that the error due to the EV approximation is negligible, and a large $T$
controls the distance between the NN and the query point. The parameter $%
\lambda $ can be any constant in the open unit interval, and hence $T$ can
be much smaller than $n$.

Under the above conditions, we establish the asymptotically correct uniform coverage
by confidence interval (\ref{S_Lambda}) in the following theorem, which is
the main result of this paper.

\begin{theorem}
\label{thm main}Suppose that Conditions 1.1-1.4 hold. For any fixed $k$ and
any $F_{Y|X=x_{0}}$ that satisfies these conditions, 
\begin{equation*}
\left. \underset{n,T\rightarrow \infty }{\lim \inf }\right. \mathbb{P}%
\left(Q_{Y|X=x_{0}}(1-h/n)\in S(\mathbf{Y})\right) \geq 1-\alpha ,
\end{equation*}
where $S\left( \cdot \right) $ is determined in (\ref{S_Lambda}).
\end{theorem}

We conclude this subsection with a discussion of main properties, advantages
and disadvantages of the our proposed fixed-$k$ confidence interval. First,
since the confidence interval is based on a fixed number $k$ of tail
observations, its length does not decrease in $n$. On the other hand, the
length decreases in $k$. Second, unlike kernel regression approaches that
require a sequence of moving tuning parameter (i.e., the bandwidth parameter tending
to zero), our method only relies on a `fixed' tuning parameter which is $k$.
While common data-driven choice rules for bandwidths are not theoretically
compatible with inference for their failure to undersmooth estimates, our
method based on any fixed $k$ of a researcher's choice guarantees asymptotically valid inference. Third, our fixed-$k$ approach allows for the confidence
interval to have a uniform size control property over a set of data generating processes involving a set of values of the tail index, while the existing methods have not been shown to share this uniformity property. 
This property of our fixed-$k$ approach is useful because $\xi$ is practically unknown to researchers
and thus the size control should be uniform for all the values of $\xi$ that
are empirically relevant.

\subsection{Extension to linear random coefficients models
\label{sec ext}}

The proof strategy for our main result, namely Theorem \ref{thm main}, and
thus our proposed method of constructing confidence intervals apply to
other contexts. Among others, inference for extremal quantiles of
random coefficients in linear regression models is also possible with
our proposed strategy. In this section, we study this class of models which
have been widely used in empirical studies in economics.

Consider the model 
\begin{equation}
Y_{it}=\alpha _{i}+X_{it}^{\intercal }\beta _{i}+u_{it},  \label{reg}
\end{equation}%
where $(\alpha _{i},\beta _{i}^{\intercal })^{\intercal }$ denotes random
coefficients and $u_{it}$ denotes an error term. This setup has been studied
by numerous papers in the literature, and covers the classic panel linear
regression model with fixed effects in which $\beta _{i}=\beta_{0}$ for all $%
i$. 
As long as Conditions 1.1-1.4 are satisfied, the previously introduced
methods naturally apply here for inference on the conditional extremal
quantiles of $Y$. In addition, the model (\ref{reg}) allows
us to conduct inference on the unconditional tail features of the random coefficients, $\alpha _{i}$
and $\beta _{i}$. The remainder of this subsection illustrates a procedure
to this end.

Let $(\hat{\alpha}_{i},\hat{\beta}_{i}^{\intercal })^{\intercal }$ be the
OLS estimator by regressing $Y_{it}$ on $(1,X_{it}^{\intercal })^{\intercal}$
using the time series associated with the $i$-th individual. Collect $\{(%
\hat{\alpha}_{i},\hat{\beta}_{i}^{\intercal })^{\intercal }\}_{i=1}^{n}$ and
sort each series of estimates in the descending order. We then define 
\begin{equation*}
\mathbf{A}=(\hat{\alpha}_{\left( 1\right) },...,\hat{\alpha}_{\left(k\right)
})^{\intercal },
\end{equation*}
that is, the largest $k$ order statistics of $\{\hat{\alpha}_{i}\}$, and 
\begin{equation*}
\mathbf{B}_{j}=(\hat{\beta}_{j,\left( 1\right) },...,\hat{\beta}%
_{j,\left(k\right) })^{\intercal },
\end{equation*}
that is, the largest $k$ order statistics of the $j$-th coordinate of $\{%
\hat{\beta}_{i}\}_{i=1}^{n}$, for each $j$. Without loss of generality, we focus on the
first coordinate of $\beta _{i}$, and suppress the subscript $j$ from our
notations for simplicity.

Now, we substitute $\mathbf{A}$ or $\mathbf{B}$ in $S(\cdot )$ as in (\ref%
{S_Lambda}) to construct the confidence interval for extremal quantiles of $%
\alpha _{i}$ or $\beta _{i}$. The following conditions are imposed for a
theoretical guarantee of correct asymptotic coverage.

\begin{description}
\item[Condition 2.1] $(\alpha _{i},\beta _{i}^{\intercal
},u_{it},X_{it}^{\intercal })^{\intercal }$ are i.i.d.\ across $i$ and
strictly stationary and weakly dependent across $t$;

\item[Condition 2.2] $F_{\alpha }\in \mathcal{D}\left( G_{\xi _{\alpha
}}\right) $ and $F_{\beta }\in \mathcal{D}\left( G_{\xi _{\beta }}\right) $
with $\xi _{\alpha } \in \Xi$ and $\xi _{\beta } \in \Xi$;

\item[Condition 2.3] $\sup_{i}\left\vert
\left\vert (\hat{\alpha}_i,\hat{\beta}_{i}^\intercal)^\intercal-(\alpha_i,\beta _{i}^\intercal)^\intercal\right\vert \right\vert=o_{p}(1)$, $%
\sup_{i}\left\vert \bar{u}_{i}\right\vert =o_{p}(1)$, and $%
\sup_{i}\left\vert \left\vert \bar{X}_{i}\right\vert \right\vert =O_{p}(1)$,
where $\bar{X}_{i}=T^{-1}\sum_{t=1}^{T}X_{it}$ and $\bar{u}%
_{i}=T^{-1}\sum_{t=1}^{T}u_{it}$. In addition, if $\xi _{\omega }=0$, $%
\sup_{i}\left\vert
\left\vert (\hat{\alpha}_i,\hat{\beta}_{i}^\intercal)^\intercal-(\alpha_i,\beta _{i}^\intercal)^\intercal\right\vert \right\vert/f_{\omega }\left( Q_{\omega
}(1-1/n)\right) \left. =\right. o_{p}(1)$ and $\sup_{i}\left\vert \bar{u}%
_{i}\right\vert /f_{\omega }\left( Q_{\omega }(1-1/n)\right) \left. =\right.
o_{p}(1)$ for $\omega =\alpha $ or $\beta $, where $Q_{\omega }(\cdot )$ and 
$f_{\omega }(\cdot )$ denote the quantile function and the PDF of $\omega $,
respectively. Alternatively, if $\xi _{\omega }<0$, $n^{-\xi _{\omega
}}T^{-1/2}\rightarrow 0$, $\sup_{i}\left\vert \left\vert \bar{X}%
_{i}\right\vert \right\vert =O_{p}\left( 1\right) $, $\sup_{i}\left\vert
\left\vert (\hat{\alpha}_i,\hat{\beta}_{i}^\intercal)^\intercal-(\alpha_i,\beta _{i}^\intercal)^\intercal\right\vert \right\vert =O_{p}\left(
T^{-1/2}\right) $ and $\sup_{i}\left\vert \bar{u}_{i}\right\vert
=O_{p}\left( T^{-1/2}\right) $.
\end{description}

Condition 2.1 is similar to Condition 1.1. Since the objects of interest are
the unconditional extremal quantiles of $\alpha _{i}$ and $\beta _{i}$, we
do not need the NN condition on covariates. The dependence structure is left
unspecified as long as it is sufficient for Condition 2.3. Condition 2.2
assumes that the distributions of $\alpha _{i}$ and $\beta _{i}$ are in the
domains of attraction of $G_{\xi _{\alpha }}$ and $G_{\xi _{\beta }}$,
respectively.
Condition 2.3 requires that the estimator $%
(\hat{\alpha}_{i},\hat{\beta}_{i}^\intercal)^\intercal$ is consistent for all $i$ and the moments of sample
averages of $u_{it}$ and $X_{it}$ across $t$ are bounded. If the tail index
is non-positive, then these bounds need to be stronger to accommodate the fact that $%
a_{n}\rightarrow 0$.\footnote{%
A straightforward calculation yields that normal distribution satisfies
Condition 2.3, if $\sup_{i}\left\vert
\left\vert (\hat{\alpha}_i,\hat{\beta}_{i}^\intercal)^\intercal-(\alpha_i,\beta _{i}^\intercal)^\intercal\right\vert \right\vert\left. =\right.
O_{p}(T^{-\varepsilon })$ and $\sup_{i}\left\vert \bar{u}_{i}\right\vert
\left. =\right. O_{p}(T^{-\varepsilon })$ for some $\varepsilon >0$ and if $%
n/T\left. \rightarrow \right. \lambda $ for some $\lambda \left. \in
\right.(0,\infty )$. This can be seen by $1/f_{\alpha }\left( Q_{\alpha
}\left( 1\left.-\right. 1/n\right) \right) \left. \leq \right. O(\log (n))$
when $f_{\alpha}$ and $Q_{\alpha }$ are standard normal density and quantile
functions, respectively (cf. Example 1.1.7 in \cite{deHaan07}).}

Under these conditions, the following corollary establishes the asymptotic
coverage.

\begin{corollary}
\label{col ab}Suppose that Conditions 1.4 and 2.1-2.3 hold. For any fixed $k$
and any $F_{\alpha }$ and $F_{\beta }$ that satisfy these conditions, 
\begin{eqnarray*}
\left. \underset{n,T\rightarrow \infty }{\lim \inf }\right. \mathbb{P}\left(
Q_{\alpha }(1-h/n)\in S(\mathbf{A})\right)  &\geq &1-\alpha  \\
\left. \underset{n,T\rightarrow \infty }{\lim \inf }\right. \mathbb{P}\left(
Q_{\beta }(1-h/n)\in S(\mathbf{B})\right)  &\geq &1-\alpha 
\end{eqnarray*}%
where $S\left( \cdot \right) $ is defined in (\ref{S_Lambda}).
\end{corollary}

\section{Monte Carlo simulation studies\label{sec MC}}

We conduct Monte Carlo experiments to examine the small sample performance
of the new approach. In Section \ref{sec MC no FE}, we first consider the
simple panel data $\{Y_{it},X_{it}\}$ without any fixed effect. In Section %
\ref{sec MC lgth}, we compare the efficiency of the new approach with the
kernel estimator, which essentially uses more than one NNs. In Section \ref%
{sec MC with RE}, we consider the linear random coefficient regression setup
(\ref{reg}).

\subsection{Conditional extremal quantiles\label{sec MC no FE}}

We continue to consider the three examples in Section \ref{sec general} as
the data generating processes (DGPs). In all experiments, generated data are
i.i.d. across $i$, but are dependent across $t$. The dependence structure across $t$ is specified as follows.

\begin{description}
\item[1. Joint Normal] $X_{it}=\rho X_{it-1}+u_{it}$ with $u_{it}\sim ^{iid}%
\mathcal{N}\left( 0,1-\rho ^{2}\right) $ and $X_{i1}\sim \mathcal{N}\left(
0,1\right) $. $Y_{it}=r_{xy}X_{it}+\sqrt{1-r_{xy}^{2}}v_{it}$ where $%
v_{it}\sim ^{iid}\mathcal{N}\left( 0,1\right) $ and independent of $u_{it}$.
Set $\rho =0.5$ and $r_{xy}=0.5.$

\item[2. Joint Student-t] $(X_{it},Y_{it})$ is i.i.d.\ across $t$ and
distributed as $t_{v}(\mu ,\Sigma )$ with $v=3$, $\mu =[0,0]^{\intercal }$,
and $\Sigma =[1,0.5;0.5,1]$.

\item[3. Conditional Pareto] $X_{it}=\rho X_{it-1}+u_{it}$ with $u_{it}\sim
^{iid}\mathcal{N}\left( 0,1-\rho ^{2}\right) $ and $X_{i1}\sim \mathcal{N}%
\left( 0,1\right) $. $Y_{it}|X_{it}=x\sim \mathrm{Pa}(\xi (x))$, that is, $%
\mathbb{P}(Y_{it}\leq y|X_{it}=x)=1-y^{-1/\xi (x)}$ for $y\geq 1$ where $\xi
\left( x\right) =x+0.5$.
\end{description}

We construct CIs for $Q_{Y|X=x_{0}}\left( 1-h/n\right) $ with $x_{0}=0$ and $%
1.65$ (the 50\% and 95\% quantiles of $X$, respectively) and $h=1$ and $5$.
The sample sizes $n$ and $T$ are either 200 or 500, with smaller
combinations exercised in later experiments.

We compare results across three approaches: (i) the fixed-$k$ approach
(fixed-$k$) introduced in this paper, (ii) quantile regression (QR), and
(iii) bootstrapping the empirical quantile (Boot). We produce the fixed-$k$
CI using $k=20$ in most cases if not otherwise noted. The space of $\xi $ is
restricted to be $[-1/2,1/2]$. For the QR approach, we run a quantile regression of $%
Y_{it}$ on $X_{it}$ and a constant at the $\tau $ quantile for each $i\,$.
The conditional quantile is estimated at $\hat{\beta}_{0i}+x_{0}\hat{\beta}%
_{1i}$ where $\hat{\beta}_{0i}$ and $\hat{\beta}_{1i}$ are the coefficient
estimates using the $i$-th individual's observations. The CI is defined the
2.5\% and 97.5\% quantiles of these $n$ estimates. The bootstrap CI is based
on bootstrapping the empirical $\tau $ quantile in $\{Y_{i,[x_{0}]}%
\}_{i=1}^{n}$. The bootstrap size is 200.

Tables \ref{tbl h5 tx05}-\ref{tbl h1 tx05} depict the coverage probabilities
(Cov) and the average lengths (Lgth) of the above three methods based on 500
simulation draws. The fixed-$k$ approach performs well in terms of both the
coverage and length across all the specifications. Regarding the QR method,
recall that the conditional quantile is a linear function of $X$ in the
first DGP but not in the other two. Therefore, not surprisingly, the CIs
based on QR perform well in the first DGP but deliver substantial
undercoverage and longer length in the other two due to misspecification.
The bootstrap approach is robust to misspecification but requires the
asymptotic normal approximation, which performs well only in the mid-sample
and does not near the tails. As such, the bootstrap intervals exhibit more
undercoverage for $h=1$ than for $h=5$.

We conclude the current subsection with a remark about the choice of $k$. A
larger $k$ leads to more tail observations and hence shorter confidence
intervals, but is subject to a larger approximation bias due to including
too many mid-sample data. This indicates that the choice of $k$ is
difficult, especially when $n$ is only moderate. It is actually impossible
to choose a uniformly best $k$ allowing the underlying CDF to be flexible
(see Theorem 1 of \cite{MuellerWang17}). The CDFs in our Monte Carlo designs
are all well behaved so that such a value of $k$ as large as 40\% of the
sample size performs well. This is seen in Table \ref{tbl h1 tx05}, which
reports the numbers for $k=20$ and $50$.

\begin{footnotesize}
\begin{table}[tbp] 
\renewcommand{\baselinestretch}{1.0}
\vspace{-1cm}
\begin{center}
\begin{footnotesize}%
\caption{Finite sample performance of inference about conditional extremal quantile,
 no model specification}\label{tbl h5 tx05}

\begin{tabular}{lccccccccccc}
\hline\hline
$n$ &  & \multicolumn{4}{c}{200 (97.5\% quantile)} &  &  & 
\multicolumn{4}{c}{500 (99\% quantile)} \\ \cline{3-12}
$T$ &  & \multicolumn{2}{c}{200} & \multicolumn{2}{c}{500} &  &  & 
\multicolumn{2}{c}{200} & \multicolumn{2}{c}{500} \\ 
&  & Cov & Lgth & Cov & Lgth &  &  & Cov & Lgth & Cov & Lgth \\ \hline
&  & \multicolumn{10}{c}{Joint Normal} \\ 
fixed-$k$ &  & 0.97 & 0.63 & 0.96 & 0.66 &  &  & 0.95 & 0.56 & 0.96 & 0.56
\\ 
QR &  & 1.00 & 0.63 & 1.00 & 0.41 &  &  & 1.00 & 0.89 & 1.00 & 0.56 \\ 
Boot &  & 0.97 & 0.64 & 0.91 & 0.61 &  &  & 0.88 & 0.58 & 0.95 & 0.55 \\ 
\hline
&  & \multicolumn{10}{c}{Joint Student-t} \\ 
fixed-$k$ &  & 0.96 & 1.35 & 0.96 & 1.47 &  &  & 0.95 & 1.62 & 0.94 & 1.63
\\ 
QR &  & 0.95 & 2.20 & 0.00 & 1.31 &  &  & 1.00 & 4.76 & 0.01 & 2.87 \\ 
Boot &  & 0.91 & 1.36 & 0.95 & 1.34 &  &  & 0.89 & 1.51 & 0.94 & 1.68 \\ 
\hline
&  & \multicolumn{10}{c}{Conditional Pareto} \\ 
fixed-$k$ &  & 0.96 & 7.65 & 0.97 & 7.14 &  &  & 0.98 & 15.8 & 0.97 & 11.6
\\ 
QR &  & 0.00 & ${>}$10$^{3}$ & 0.00 & ${>}$10$^{3}$ &  &  & 0.00 & ${>}$10$%
^{3}$ & 0.00 & ${>}$10$^{3}$ \\ 
Boot &  & 0.93 & 8.30 & 0.93 & 7.80 &  &  & 0.94 & 15.3 & 0.90 & 12.7 \\ 
\hline
\end{tabular}

\end{footnotesize}
\end{center}
\renewcommand{\baselinestretch}{0.9} \normalsize \footnotesize%
Note: Entries are coverages and lengths of the CIs for $Q_{Y|X=0}(1-5/n)$.
See the main text for the description of the three approaches and the data
generating processes. Confidence level is 5\%. Based on 500 simulation draws.%
\end{table}%
\end{footnotesize}%

\begin{footnotesize}
\begin{table}[tbp] 
\renewcommand{\baselinestretch}{1.0}
\vspace{-1cm}
\begin{center}
\begin{footnotesize}%
\caption{Finite sample performance of inference about conditional extremal quantile,
 no model specification}\label{tbl h5 tx095}

\begin{tabular}{lccccccccccc}
\hline\hline
$n$ &  & \multicolumn{4}{c}{200 (97.5\% quantile)} &  &  & 
\multicolumn{4}{c}{500 (99\% quantile)} \\ \cline{3-12}
$T$ &  & \multicolumn{2}{c}{200} & \multicolumn{2}{c}{500} &  &  & 
\multicolumn{2}{c}{200} & \multicolumn{2}{c}{500} \\ 
&  & Cov & Lgth & Cov & Lgth &  &  & Cov & Lgth & Cov & Lgth \\ \hline
&  & \multicolumn{10}{c}{Joint Normal} \\ 
fixed-$k$ &  & 0.96 & 0.65 & 0.96 & 0.65 &  &  & 0.95 & 0.57 & 0.94 & 0.57
\\ 
QR &  & 1.00 & 1.28 & 1.00 & 0.80 &  &  & 1.00 & 1.81 & 1.00 & 1.13 \\ 
Boot &  & 0.93 & 0.63 & 0.92 & 0.64 &  &  & 0.92 & 0.55 & 0.91 & 0.56 \\ 
\hline
&  & \multicolumn{10}{c}{Joint Student-t} \\ 
fixed-$k$ &  & 0.97 & 2.42 & 0.95 & 2.36 &  &  & 0.97 & 2.88 & 0.97 & 2.77
\\ 
QR &  & 1.00 & 3.53 & 1.00 & 2.26 &  &  & 1.00 & 6.23 & 1.00 & 3.94 \\ 
Boot &  & 0.94 & 2.31 & 0.93 & 2.25 &  &  & 0.93 & 2.83 & 0.95 & 2.72 \\ 
\hline
&  & \multicolumn{10}{c}{Conditional Pareto} \\ 
fixed-$k$ &  & 0.95 & 9.30 & 0.97 & 7.45 &  &  & 0.84 & 16.3 & 0.94 & 12.5
\\ 
QR &  & 0.00 & ${>}$10$^{3}$ & 0.00 & ${>}$10$^{3}$ &  &  & 0.00 & ${>}$10$%
^{3}$ & 0.00 & ${>}$10$^{3}$ \\ 
Boot &  & 0.95 & 12.5 & 0.96 & 8.91 &  &  & 0.80 & 26.8 & 0.93 & 15.2 \\ 
\hline
\end{tabular}

\end{footnotesize}
\end{center}
\renewcommand{\baselinestretch}{0.9} \normalsize \footnotesize%
{\footnotesize Note: Entries are coverages and lengths of the CIs for }$%
Q_{Y|X=1.65}(1-5/n)${\footnotesize . See the main text for the description
of the three approaches and the data generating processes. Confidence level
is 5\%. Based on 500 simulation draws.}%
\end{table}%
\end{footnotesize}%

\begin{footnotesize}
\begin{table}[tbp] 
\renewcommand{\baselinestretch}{1.0}
\vspace{-1cm}
\begin{center}
\begin{footnotesize}%
\caption{Finite sample performance of inference about conditional extremal quantile,
 no model specification}\label{tbl h1 tx05}

\begin{tabular}{lccccccccccc}
\hline\hline
$n$ &  & \multicolumn{4}{c}{200 (99.5\% quantile)} &  &  & 
\multicolumn{4}{c}{500 (99.8\% quantile)} \\ \cline{3-12}
$T$ &  & \multicolumn{2}{c}{200} & \multicolumn{2}{c}{500} &  &  & 
\multicolumn{2}{c}{200} & \multicolumn{2}{c}{500} \\ 
&  & Cov & Lgth & Cov & Lgth &  &  & Cov & Lgth & Cov & Lgth \\ \hline
&  & \multicolumn{10}{c}{Joint Normal} \\ 
fixed-$k$(k=20) &  & 0.95 & 1.82 & 0.96 & 1.83 &  &  & 0.97 & 1.69 & 0.96 & 
1.70 \\ 
QR &  & 1.00 & 1.19 & 1.00 & 0.75 &  &  & 1.00 & 1.18 & 1.00 & 1.07 \\ 
Boot &  & 0.63 & 0.62 & 0.64 & 0.59 &  &  & 0.64 & 0.57 & 0.65 & 0.59 \\ 
\hline
&  & \multicolumn{10}{c}{Joint Student-t} \\ 
fixed-$k$(k=20) &  & 0.96 & 4.71 & 0.96 & 4.69 &  &  & 0.96 & 5.62 & 0.97 & 
5.61 \\ 
fixed-$k$(k=50) &  & 0.94 & 3.91 & 0.92 & 3.90 &  &  & 0.95 & 4.85 & 0.92 & 
4.73 \\ 
QR &  & 1.00 & 8.51 & 0.68 & 5.51 &  &  & 1.00 & 8.47 & 1.00 & 11.5 \\ 
Boot &  & 0.62 & 2.01 & 0.60 & 2.02 &  &  & 0.63 & 2.57 & 0.61 & 2.56 \\ 
\hline
&  & \multicolumn{10}{c}{Conditional Pareto} \\ 
fixed-$k$(k=20) &  & 0.98 & 27.6 & 0.98 & 26.1 &  &  & 0.94 & 48.1 & 0.97 & 
40.5 \\ 
QR &  & 0.00 & ${>}$10$^{3}$ & 0.00 & ${>}$10$^{3}$ &  &  & 0.00 & ${>}$10$%
^{3}$ & 0.00 & ${>}$10$^{3}$ \\ 
Boot &  & 0.71 & 25.9 & 0.63 & 30.4 &  &  & 0.78 & 76.2 & 0.77 & 43.9 \\ 
\hline
\end{tabular}

\end{footnotesize}
\end{center}
\renewcommand{\baselinestretch}{0.9} \normalsize \footnotesize%
Note: Entries are coverages and lengths of the CIs for $Q_{Y|X=0}(1-1/n)$.
See the main text for the description of the three approaches and the data
generating processes. Confidence level is 5\%. Based on 500 simulation draws.%
\end{table}%
\end{footnotesize}%

\subsection{Comparison with kernel smoothing\label{sec MC lgth}}

Our new approach takes only one NN in each time series, which raises the
question of efficiency loss. We answer this by comparing our fixed-$k$
approach with the kernel smoothing method proposed by \cite{Gardes10}. In
particular, we first pool the panel data into a cross-sectional sample.
Suppose the object of interest is still $Q_{Y|X=x_{0}}(\tau )$. We follow 
\cite{Gardes10} to pick the bin $B_{b_{nT}}(x_{0})$ centered at $x_{0}$ with
a bandwidth $b_{nT}$. Since there is no theoretical justification for the
optimal choice of $b_{nT}$, we take the rule-of-thumb choice $c(nT)^{-1/5}$
with different values of the constant $c$. Now a certain choice of $b_{nT}$
leads to a certain collection of $Y^{\prime }s$ whose paired $X^{\prime }s$
are in the bin $B_{b_{nT}}(x_{0})$. Sort these induced $Y^{\prime }s$ in the
descending order into $\{Y_{(1)}\geq Y_{(2)}\geq ...\geq Y_{(m)}\}$ where $m$
denotes the local sample size determined by the bandwidth. Such local sample
size is approximately $nTb_{n}$ in the kernel smoothing (as opposed to $n$
in our new approach).

Given the induced $Y$'s, the conditional quantile is estimated as $\hat{Q}%
_{Y|X=x_{0}}(\tau ) {\footnotesize =} Y_{(\left\lfloor (1-\tau
)m\right\rfloor )}$, that is, the $\left\lfloor (1-\tau )m\right\rfloor $-th
largest order statistics in the induced $Y$'s where $\left\lfloor (1-\tau
)m\right\rfloor $ denotes the integer part of $\left( 1-\tau \right) m$. 
\cite{Gardes10} show that, under $m(1-\tau )\rightarrow \infty $ and some
other regularity conditions, 
\begin{equation*}
\sqrt{m(1-\tau )}\left( \frac{{\footnotesize \hat{Q}}_{{\footnotesize Y|X=}%
x_{0}}{\footnotesize (\tau )}}{{\footnotesize Q}_{{\footnotesize Y|X=}x_{0}}%
{\footnotesize (\tau )}}-1\right) \overset{d}{\rightarrow }\mathcal{N}%
(0,1/\xi _{0}^{2}(x_{0})).
\end{equation*}
Then, the CI of $Q_{{\footnotesize Y|X=}x_{0}}(\tau )$ is constructed by the
delta method and plugging in some consistent estimator of $\xi _{0}$. One
choice which they propose is the Hill-type estimator 
\begin{equation}
1/\hat{\xi}=\frac{1}{k-1}\sum_{i=1}^{k-1}i\log (Y_{(i)}/Y_{(i+1)})
\label{Smooth Hill}
\end{equation}%
for some choice of $k<m$.

For comparisons, we implement our fixed-$k$ approach by using the panel data
and the above kernel approach by pooling the data. In particular, we
implement the conditional Pareto DGP in the previous experiment with $n=200$
and $T$ ranging from 50 to 500. For the fixed-$k$ CI, we set $k=50$. For the
kernel method, we implement $c\in \{0.1,0.25,0.5,1,2\}$ and set $k$ (in the
Hill-type index estimator (\ref{Smooth Hill})) as the largest integer less
than or equal to $m/4$.

Table \ref{tbl kernel one x} presents the coverages and the lengths of the
fixed-$k$ and the kernel CIs. Several interesting observations can be made.
First, the kernel approach is sensitive to the choice of the bandwidth. In
particular, a correct coverage relies on a narrow window of the bandwidth
choice. A larger choice can lead to a substantial undercoverage since the
smoothing bias dominates quickly in the tail. Second, when $T$ is only
moderately large (say 25 and 50), the fixed-$k$ CIs are much shorter than
the kernel one and both of them have good coverage properties. This is
because the fully nonparametric method ignores the domain-of-attraction
information, which is utilized by our fixed-$k$ method. Third, when $T$ is
very large, say 500, choosing only one NN does incur an efficiency loss as
we compare the lengths between our fixed-$k$ and the kernel CIs. But such a
loss is approximately in a factor of two or three instead of $T^{1/2}$. This
means that a general covariate-dependent tail is very difficult to estimate
in a fully nonparametric way.

%
\begin{footnotesize}
\begin{table}[tbp] 
\renewcommand{\baselinestretch}{1.0}

\begin{center}
\begin{footnotesize}%
\caption{Finite sample performance of inference about conditional extremal quantile,
 comparison with kernel method}\label{tbl kernel one x}

\begin{tabular}{lccccccccc}
\hline
$T$ &  & \multicolumn{2}{c}{50} & \multicolumn{2}{c}{100} & 
\multicolumn{2}{c}{200} & \multicolumn{2}{c}{500} \\ 
&  & Cov & Lgth & Cov & Lgth & Cov & Lgth & Cov & Lgth \\ \hline
fixed-k &  & 0.97 & 21.1 & 0.97 & 19.8 & 0.97 & 16.8 & 0.98 & 15.3 \\ 
NP(c=0.1) &  & 0.91 & 50.1 & 0.89 & 30.0 & 0.94 & 24.4 & 0.93 & 14.6 \\ 
NP(c=0.25) &  & 0.94 & 33.1 & 0.96 & 19.0 & 0.93 & 13.3 & 0.96 & 9.15 \\ 
NP(c=0.5) &  & 0.93 & 17.1 & 0.94 & 12.7 & 0.93 & 9.28 & 0.95 & 6.36 \\ 
NP(c=1) &  & 0.93 & 13.9 & 0.90 & 9.84 & 0.89 & 7.14 & 0.88 & 4.77 \\ 
NP(c=2) &  & 0.37 & 15.6 & 0.24 & 10.1 & 0.14 & 6.86 & 0.11 & 4.18 \\ \hline
\end{tabular}

\end{footnotesize}
\end{center}
\renewcommand{\baselinestretch}{0.9} \normalsize \footnotesize%
Note: Entries are coverages and lengths of the CIs for $Q_{Y|X=0}(1-1/n)$
under the conditional Pareto DGP. See the main text for the description of
the two approaches and details of the DGP. Confidence level is 5\%. Based on
500 simulation draws.%
\end{table}%
\end{footnotesize}%

In Table \ref{tbl kernel two x}, we consider a two-dimensional standard
normal $X$ and generate $Y_{it}$ by $Y_{it}|X_{it}=x\sim \pm $Pa$(\xi (x))$
with $\xi (x_{1},x_{2})=x_{1}+x_{2}+0.5$. The kernel method is illustrated
with $c\in \{0.5,1,2,4\}$. All the other parameter choices for both methods
remain unchanged as those used for Table \ref{tbl kernel one x}. The results
clearly suggest that our fixed-$k$ method together with the NN choice
dominates the kernel method in terms of both the coverage probabilities and
lengths. In particular, the kernel method suffers from the curse of
dimensionality as the dimension of $X$ increases.

\begin{footnotesize}
\begin{table}[tbp] 
\renewcommand{\baselinestretch}{1.0}
\begin{center}
\begin{footnotesize}%
\caption{Finite sample performance of inference about conditional extremal quantile,
 comparison with kernel method, two-dimensional X}\label{tbl kernel two x}

\begin{tabular}{lccccccccc}
\hline
$T$ &  & \multicolumn{2}{c}{50} & \multicolumn{2}{c}{100} & 
\multicolumn{2}{c}{200} & \multicolumn{2}{c}{500} \\ 
&  & Cov & Lgth & Cov & Lgth & Cov & Lgth & Cov & Lgth \\ \hline
fixed-k &  & 0.96 & 21.0 & 0.97 & 17.2 & 0.96 & 16.5 & 0.96 & 16.3 \\ 
NP(c=0.5) &  & 0.56 & 65.8 & 0.73 & 60.6 & 0.73 & 54.3 & 0.81 & 53.8 \\ 
NP(c=1) &  & 0.83 & 75.0 & 0.80 & 60.8 & 0.93 & 60.9 & 0.91 & 32.0 \\ 
NP(c=2) &  & 0.96 & 66.1 & 1.00 & 36.0 & 0.97 & 25.2 & 0.97 & 16.5 \\ 
NP(c=4) &  & 0.74 & 54.5 & 0.47 & 35.4 & 0.23 & 22.8 & 0.16 & 13.2 \\ \hline
\end{tabular}

\end{footnotesize}
\end{center}
\renewcommand{\baselinestretch}{0.9} \normalsize \footnotesize%
Note: Entries are coverages and lengths of the CIs for $Q_{Y|X=0}(1-1/n)$
under the conditional Pareto DGP. See the main text for the description of
the two approaches and details of the DGP. Confidence level is 5\%. Based on
500 simulation draws. 
\end{table}%
\end{footnotesize}%

As a final remark of this subsection, we also implement the standard kernel
weighted quantile regression method designed for the mid-sample quantiles
(cf.\ Chapter 10 of \cite{LiRacine07}). Given a large $T$, the target $1-1/n$
conditional quantile is relatively in the mid-sample after pooling the panel
data into a cross-sectional one, and hence the confidence interval based on
asymptotic normality might work. However, unreported Monte Carlo simulations
show that this method works only if $T$ is substantially larger than (e.g,
five times as much as) $n$. In our experiments, it is strictly dominated by
the method proposed by \cite{Gardes10}.

\subsection{Extremal quantiles in a linear random coefficient model\label%
{sec MC with RE}}

In this section, we consider the linear random coefficient model $%
Y_{it}=\alpha _{i}+X_{it}\beta _{0}+u_{it}$, where the generated
observations including the random coefficients are i.i.d. across $i$. For
the time series dependence, we set $\alpha_{i}=T^{-1}\sum_{t=1}^{T}X_{it}$
and $X_{it}=\rho X_{it-1}+e_{it}$ with $e_{it}\sim ^{iid}\mathcal{N}\left(
0,(1-\rho ^{2})\right) $ and $X_{i0}\sim \mathcal{N}\left( 0,1\right) $. The
conditional distributions of $u_{it}$ given $X_{it}=x$ are specified as
follows.

\begin{description}
\item[1. Conditional Normal] $u_{it}|X_{it}=x\sim \mathcal{N}\left(
0,1+x^{2}\right) $.

\item[2. Conditional Student-t] $u_{it}|X_{it}=x\sim t\left( 2+\left\vert
x\right\vert \right) $.

\item[3. Conditional Pareto] $u_{it}|X_{it}=x\sim \pm \mathrm{Pa}(\xi (x))$,
that is, $\mathbb{P}(u_{it}\leq y|X_{it}=x)=1/2+(1-(1+y)^{-1/\xi (x)})/2$
for $y\geq 0$, and $\mathbb{P}(u_{it}\leq y|X_{it}=x)=\left( -y+1\right)
^{-1/\xi (x)}/2$ for $y\leq 0$ where $\xi \left( x\right) =x+0.5$.
\end{description}

We use the same set of the three approaches as in the Section \ref{sec MC no
FE} to construct CIs for the conditional extremal quantile $%
Q_{Y_{it}|X_{it}=x_{0}}\left( \tau \right) =Q_{\varepsilon
_{it}|X_{it}=x_{0}}\left( \tau \right) +x_{0}\beta _{0}$, where $\varepsilon
_{it}$ denotes $\alpha _{i}+u_{it}$. Specifically, our fixed-$k$ approach is
conducted in two ways: with or without using the standard within least
squares estimator of $\beta _{0}$. For the former (fixed-$k$ w.\thinspace
LS), we first estimate $\beta _{0}$ using the standard within estimator $%
\hat{\beta}$ and back out $\hat{\varepsilon}_{it}=Y_{it}-X_{it}\hat{\beta}$.
We then implement the steps in Section \ref{sec general} to construct the
CIs for the conditional quantiles of $\varepsilon _{it}$. The CIs for $%
Q_{Y_{it}|X_{it}=x_{0}}\left( \tau \right) $ are obtained by adding back $%
x_{0}\hat{\beta}$. For the one ignoring the linear regression structure
(fixed-$k$ w/o LS), we directly use $(Y_{it},X_{it})^\intercal$, and apply Steps 1-3
in Section \ref{sec general}.

Table \ref{tbl h1 tx05 RE} presents the results for $n\in \{100,200\}$ and $%
T\in \{25,50,200,500\}$. Several interesting observations can be made.
First, the errors in the conditional t and conditional Pareto models do not
have finite variances when $x_{0}$ is 0, and hence the LS estimator of $%
\beta _{0}$ behaves poorly. This leads to a poor performance of the fixed-$k$
approach if the linear regression model is utilized. This problem can be
solved by using the least absolute deviation (LAD) estimator as shown in
unreported results. In comparison, the fixed-$k$ CIs without using the
linear regression model always perform well given a large enough sample
size. Second, the QR approach still suffers from undercoverage in all three
specifications since the normal and the Student-t DGPs have nonlinear
heteroskedasticity and the conditional Pareto DGP violates the constant tail
shape condition. Finally, the bootstrap method performs poorly if the
extremal quantiles under investigation are too far in the tail.

\begin{footnotesize}
\begin{table}[tbp] 
\renewcommand{\baselinestretch}{1.0}
\vspace{-1cm}
\begin{center}
\begin{footnotesize}%
\caption{Finite sample performance of inference about conditional extremal quantile,
 non-dynamic model with random effects}\label{tbl h1 tx05 RE}

\begin{tabular}{lccccccccccc}
\hline\hline
$n$ &  & \multicolumn{4}{c}{200 (99.5\% quantile)} &  &  & 
\multicolumn{4}{c}{100 (99\% quantile)} \\ \cline{3-12}
$T$ &  & \multicolumn{2}{c}{200} & \multicolumn{2}{c}{500} &  &  & 
\multicolumn{2}{c}{25} & \multicolumn{2}{c}{50} \\ 
&  & Cov & Lgth & Cov & Lgth &  &  & Cov & Lgth & Cov & Lgth \\ \hline
&  & \multicolumn{10}{c}{Conditional Normal} \\ 
fixed-$k$ w. LS &  & 0.94 & 2.23 & 0.93 & 2.13 &  &  & 0.80 & 2.85 & 0.88 & 
2.34 \\ 
fixed-$k$ w/o LS &  & 0.93 & 2.22 & 0.92 & 2.14 &  &  & 0.53 & 3.17 & 0.76 & 
2.62 \\ 
QR &  & 0.00 & 3.04 & 0.00 & 2.19 &  &  & 1.00 & 3.10 & 1.00 & 2.96 \\ 
Boot &  & 0.73 & 0.77 & 0.67 & 0.71 &  &  & 0.73 & 1.11 & 0.81 & 0.92 \\ 
\hline
& \multicolumn{11}{c}{Conditional Student-t} \\ 
fixed-$k$ w. LS &  & 0.95 & 15.3 & 0.94 & 15.5 &  &  & 0.93 & 10.0 & 0.96 & 
10.4 \\ 
fixed-$k$ w/o LS &  & 0.95 & 15.3 & 0.94 & 15.5 &  &  & 0.91 & 10.0 & 0.93 & 
10.3 \\ 
QR &  & 1.00 & 26.5 & 1.00 & 7.98 &  &  & 0.99 & 11.0 & 1.00 & 15.0 \\ 
Boot &  & 0.56 & 11.5 & 0.60 & 11.0 &  &  & 0.47 & 6.32 & 0.51 & 6.01 \\ 
\hline
& \multicolumn{11}{c}{Conditional Pareto} \\ 
fixed-$k$ w. LS &  & 0.00 & 16.2 & 0.00 & 5.90 &  &  & 0.02 & 59.6 & 0.01 & 
58.1 \\ 
fixed-$k$ w/o LS &  & 0.97 & 18.8 & 0.97 & 16.9 &  &  & 0.95 & 16.0 & 0.96 & 
15.4 \\ 
QR &  & 0.00 & ${>}$10$^{3}$ & 0.00 & ${>}$10$^{3}$ &  &  & 1.00 & ${>}$10$%
^{3}$ & 1.00 & ${>}$10$^{3}$ \\ 
Boot &  & 0.71 & 16.2 & 0.67 & 16.8 &  &  & 0.76 & 313 & 0.78 & 31.4 \\ 
\hline
\end{tabular}

\end{footnotesize}
\end{center}
\renewcommand{\baselinestretch}{0.9} \normalsize \footnotesize%
Note: Entries are coverages and lengths of the CIs for $Q_{Y|X=0}(1-1/n)$.
See the main text for the description of different approaches and the data
generating processes. Confidence level is 5\%. Based on 500 simulation draws.%
\end{table}%
\end{footnotesize}%

In Table \ref{tbl a and b}, we study the CIs for high quantiles of $%
\alpha_{i}$ and $\beta _{i}$ with data generated from $Y_{it}=%
\alpha_{i}+X_{it}\beta _{i}+u_{it}$, where $\left( \alpha _{i},\beta
_{i},X_{it},u_{it}\right) ^{\intercal }\sim ^{iid}\mathcal{N}%
\left(0,I_{4}\right) $. The i.i.d. condition is across both $i$ and $t$ in
this setting. We first estimate $\alpha _{i}$ and $\beta _{i}$ by regressing 
$Y_{it}$ on $(1,X_{it})^{\intercal }$ with $T$ observations from individual $%
i$. We then collect the estimates, $\hat{\alpha}_{i}$ and $\hat{\beta}_{i}$,
for all $i$ and sort them in the descending order to apply each of the fixed-%
$k$, QR, and bootstrap methods. The QR estimator is simply the empirical
quantile among the estimators for all $i$, whose asymptotic variance is
estimated by the standard kernel density estimator with the rule-of-thumb
bandwidth. The results suggest that the fixed-$k$ approach with NN dominates
the other two in both coverage and length, especially when the sample size
is only moderate.

\begin{footnotesize}
\begin{table}[tbp] 
\renewcommand{\baselinestretch}{1.0}
\vspace{-1cm}
\begin{center}
\begin{footnotesize}%
\caption{Finite sample performance of inference about large quantiles of the
random coefficients}\label{tbl a and b}

\begin{tabular}{lccccccccccc}
\hline\hline
$n$ &  & \multicolumn{4}{c}{$200$} &  &  & \multicolumn{4}{c}{$500$} \\ \cline{3-12}
$T$ &  & \multicolumn{2}{c}{$10$} & \multicolumn{2}{c}{$20$} &  &  & 
\multicolumn{2}{c}{$10$} & \multicolumn{2}{c}{$20$} \\ 
&  & Cov & Lgth & Cov & Lgth &  &  & Cov & Lgth & Cov & Lgth \\ \hline
&  & \multicolumn{10}{c}{CIs for $Q_{\alpha }(1-5/n)$} \\ 
fixed-$k$ &  & 0.92 & 0.77 & 0.95 & 0.76 &  &  & 0.92 & 0.69 & 0.93 & 0.67
\\ 
QR &  & 0.84 & 1.13 & 0.90 & 1.07 &  &  & 0.88 & 2.96 & 0.94 & 2.94 \\ 
Boot &  & 0.89 & 0.81 & 0.93 & 0.76 &  &  & 0.81 & 0.70 & 0.91 & 0.69 \\ 
\hline
&  & \multicolumn{10}{c}{CIs for $Q_{\beta }(1-5/n)$} \\ 
fixed-$k$ &  & 0.91 & 0.81 & 0.96 & 0.76 &  &  & 0.86 & 0.69 & 0.96 & 0.67
\\ 
QR &  & 0.81 & 1.15 & 0.88 & 1.08 &  &  & 0.88 & 3.21 & 0.94 & 2.83 \\ 
Boot &  & 0.85 & 0.82 & 0.92 & 0.76 &  &  & 0.78 & 0.73 & 0.91 & 0.68 \\ 
\hline
&  & \multicolumn{10}{c}{CIs for $Q_{\alpha }(1-1/n)$} \\ 
fixed-$k$ &  & 0.91 & 2.32 & 0.93 & 2.13 &  &  & 0.87 & 2.10 & 0.94 & 1.96
\\ 
QR &  & 0.89 & 1.45 & 0.91 & 1.42 &  &  & 0.89 & 1.33 & 0.88 & 1.29 \\ 
Boot &  & 0.57 & 0.52 & 0.58 & 0.51 &  &  & 0.57 & 0.47 & 0.54 & 0.47 \\ 
\hline
&  & \multicolumn{10}{c}{CIs for $Q_{\beta }(1-1/n)$} \\ 
fixed-$k$ &  & 0.88 & 2.32 & 0.94 & 2.28 &  &  & 0.85 & 2.16 & 0.93 & 1.93
\\ 
QR &  & 0.90 & 1.52 & 0.91 & 1.52 &  &  & 0.86 & 1.41 & 0.88 & 1.29 \\ 
Boot &  & 0.57 & 0.55 & 0.58 & 0.54 &  &  & 0.55 & 0.51 & 0.58 & 0.45 \\ 
\hline
\end{tabular}

\end{footnotesize}
\end{center}
\renewcommand{\baselinestretch}{0.9} \normalsize \footnotesize%
Note: The entries are coverage and length of the confidence intervals based
on (i) the fixed-$k$ approach using the largest k=20$\ $estimated
coefficients, (ii) empirical quantile of the estimated coefficients with
asymptotic normal approximation, and (iii) empirical quantile function of
the estimated coefficients and bootstrap. Data are generated from $%
Y_{it}=\alpha _{i}+X_{it}\beta _{i}+u_{it}$ where $\left( \alpha _{i},\beta
_{i},X_{it},u_{it}\right) ^{\intercal }\sim ^{iid}\mathcal{N}\left(
0,I_{4}\right) $. The target is the 1-h/n quantile of $\alpha _{i}$ and $%
\beta _{i}$ with $h=1$ and $5$, corresponding to 97.5\%, 98\%, 99\%, and
99.8\% quantiles\ given $n=200$\ and $500$, respectively. Confidence level
is 5\%. Based on 500 simulation draws.%
\end{table}%
\end{footnotesize}%

\section{Empirical application to extremal birth weights\label%
{sec:application}}

In this section, we reconsider the extremely low birth weights and their
relationships with mother's demographic characteristics and maternal
behaviors, which addresses an important question in health economics. We use
the detailed natality data published by the National Center for Health
Statistics, which has been used by \cite{Abrevaya01}, \cite{Koenker01}, and 
\cite{Chernozhukov11} among many others. We follow these preceding studies,
but our analysis is different from theirs in two aspects. First, these
preceding studies use the cross-sectional data in one time period, while we
collect the repeated cross-sectional samples from January 1989 to December
2002.\footnote{%
We chose this specific period for two reasons. First, this period contains
the time period of the cross-sectional data used by \cite{Abrevaya01}, \cite%
{Koenker01}, and \cite{Chernozhukov11}. Second, these periods maintain the
identical variable definitions.} Second, the previous studies all made some
parametric model assumptions, including either the linear projection model
or the (extremal) quantile regression model. In contrast, our fixed-$k$
method is nonparametric, allowing for nonparametric joint distributions.
Accordingly, some of our findings are different from those in the previous
studies.

\begin{figure}[tbp]
\caption{Plot of confidence intervals for the conditional extremal quantile
of infant birth weight.}\label{fig natality} \vspace{-4ex}
\par
\begin{center}
\includegraphics[width=0.85%
\textwidth]{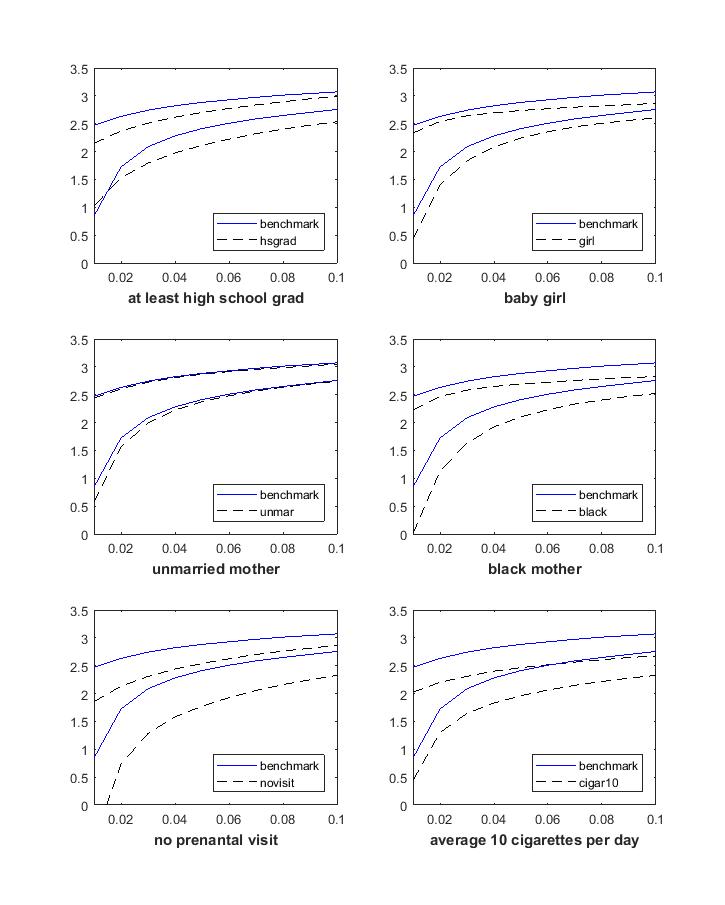}
\end{center}
\par
\vspace{-4ex}
\par
{\footnotesize Note: This figure plots the 95\% fixed-$k$ confidence
intervals for the conditional $p$-quantile of infant birth weight with $p\in
\lbrack 0.01,0.1]$, conditional on mother's age being 27, net weight gain
during pregnancy being 30 pounds, and the other six discrete covarietes. See
the main text for more detailed descriptions of these six covariates. The
vertical axis is the birth weight in kilograms, and the horizontal axis is $p
$. Data are available at the National Center for Health Statistics:
https://www.cdc.gov/nchs/nvss/births.htm. }
\end{figure}

Details of our implementation are as follows. First, we follow the
previously mentioned literature -- \cite{Abrevaya01} in particular -- to choose included covariates. Our dependent
variable is the infant birth weight measured in kilograms, and the
continuous covariates include mother's age and net weight gain (wtgain)
during pregnancy. All the remaining covariates are discrete, and hence we
consider the subsamples constructed from various combinations of the
categorical variables. For comparison, we set a benchmark subsample in which
the infant is a boy, the mother is white and married, has levels of
education less than a high school degree, had her first prenatal visit in
the first trimester (natal1), and did not smoke during pregnancy. Second,
since the samples are repeated cross-sectional, it is more natural to switch
the labeling of the indices $i$ and $t$, and first take the NN within each month. The
query point is set at age equal to 27 and wtgain equal to 30, corresponding
to their respective median values. The NN is then measured by the Euclidean
norm after standardizing each of the two variables with mean zero and unit variance. Using the same notation as in Section \ref{sec main}, we have $n=168$
and $T$ is at least 100 in every subsample. Thus, our fixed-$k$ asymptotic
framework with a large $n$ and a large $T$ is suitable with this data.
Third, we set $k=30$ based on our simulation results in the previous section
and construct the 95\% fixed-$k$ confidence intervals for the conditional $p$%
-quantiles with $p$ ranging from 1\% to 10\%. Figure \ref{fig natality}
depicts these confidence intervals in the benchmark subsample and six
alternative subsamples corresponding to one and only one of following
scenarios: the mother has at least high school diploma; the infant is a
girl; the mother is unmarried; the mother is black; the mother does not have
prenatal visit during pregnancy; and the mother smokes 10 cigarettes per day
on average.\footnote{%
For the majority of the subsamples, the number of cigarettes as recorded in
data takes only a few discrete values, including 0, 5, 10, and 20.
Therefore, we treat it as a discrete random variable in our study.}

We can make the following observations in Figure \ref{fig natality}. First,
the effects of changing the covariates are found to have a similar pattern
as in the previous studies. In particular, compared with the benchmark
subsample, the conditional quantile of infant birth weight decreases
substantially if the mother is black, did not have a prenatal visit, and/or
smoked during pregnancy. These results reconfirm the signs of the effects reported by the previous
studies. Second, on the other hand, the magnitudes of these effects are larger than
those documented in the previous studies. Specifically, \cite{Abrevaya01}
finds that smoking ten cigarettes leads to approximately 200 fewer grams at
the 10th percentile of the infant birth weight, compared with smoking no
cigarettes. \cite{Chernozhukov11} finds the quantile regression coefficient
associated with the number of cigarettes is nearly zero at the 1st
percentile (their Figure 8). On the other hand, the last sub-figure in
Figure \ref{fig natality} suggests that the difference can be over 1000
grams at the 1st percentile if we compare the mid-value between the upper
and lower bounds of the confidence intervals between these two subsamples.
Finally, the effects on extremal birth weight quantiles induced by the
demographic characteristics vary across levels of quantiles, instead of
remaining fixed.

\section{Concluding remarks\label{sec conclusion}}

This paper develops a new nonparametric method of inference for conditional
extremal quantiles using a fixed number $k$ of nearest-neighbor tail
observations in repeated cross-sectional or panel data. 
There are three advantages of our proposed method.
First, it is robust against flexible distributional assumptions unlike parametric methods.
Second, the procedure yields asymptotically valid confidence intervals for any fixed tuning parameter $k$, unlike existing kernel methods that rely on a sequence of moving tuning parameters for asymptotically valid inference.
Third, our confidence intervals enjoy the uniform coverage property over a set of data generating processes involving a set of values of the tail index.

The key insight is that the induced order statistics in each time series can be treated as
approximately stemming from the true conditional distribution, and the large
order statistics among these induced values can then be used to make
inference on extremal quantiles. By focusing on the induced order
statistics, we effectively reduce the conditional tail problem into an
unconditional one. Monte Carlo simulations show that the new method delivers
preferred small sample performance in terms of coverage probability and
length.

The new method is more flexible than the extremal quantile regression
because the latter assumes that the conditional extremal quantile is a
parametric location-shift model. If a linear regression model is imposed, then our proposed method can be easily combined with any existing consistent estimator of
structural parameters and applies to inference on extremal quantiles of the
random coefficients.

Applying the proposed method to Natality Vital Statistics, we reexamine
factors of extremely low birth weights that have been analyzed by preceding
studies. We find that signs of major effects are the same as those found in
preceding studies based on parametric models. On the other hand, we find
that some effects exhibit different magnitudes from those reported in the
previous studies based on parametric models.

\appendix
\renewcommand{\baselinestretch}{1.2} {\normalsize \sloppy
\allowdisplaybreaks
}

\section{ Appendix}

{\small This appendix provides the proof of Theorem \ref{thm main}, some
computational details, and discussions about some primitive conditions. }

\subsection{ Proofs}

{\small \label{sec:proof} }

{\small To establish Theorem \ref{thm main}, we first establish the
following intermediate result, which establishes the rate of convergence fo
the NN\ to the query point. }

\begin{lemma}
{\small \label{lemma NN}Under Condition 1.1, for each $i$ and for some $\eta
>0$, 
\begin{eqnarray}
\left\vert \left\vert X_{i,\left( x_{0}\right) }-x_{0}\right\vert
\right\vert &=&o_{a.s.}(T^{-\eta })\text{ and }  \label{uniform o1} \\
\mathbb{E}\left[ \left\vert \left\vert X_{i,\left( x_{0}\right)
}-x_{0}\right\vert \right\vert \right] &=&O\left( T^{-1/2}\right) .
\label{mse o1}
\end{eqnarray}
}
\end{lemma}

\paragraph{{ Proof of Lemma \protect\ref{lemma NN}}}

{\small We first prove (\ref{uniform o1}). The subscript $i$ is suppressed
for notional ease. Define $D_{t}\left. =\right. ||X_{t}-x_{0}||$ for $t\in
\{1,\ldots ,T\}$, which is still strictly stationary and $\beta $-mixing. By
Berbee's lemma (enlarging the probability space as necessary), the process $%
\{D_{t}\}$ can be coupled with a process $\{D_{t}^{\ast }\}$ that satisfies
the following three properties: (i) $Z_{i}\equiv \{D_{(i-1)\times
q_{T}+1},\ldots ,D_{i\times q_{T}}\}$ and $Z_{i}^{\ast }\equiv
\{D_{(i-1)\times q_{T}+1}^{\ast },\ldots ,D_{i\times q_{T}}^{\ast }\}$ are
identically distributed for all $i\left. \in \right. \{1,\ldots ,k_{T}\}$,
where $Z_{i}^{\ast }$ is the same decomposition of $\{D_{t}^{\ast }\}$ as $%
Z_{i}$ and $k_{T}\left. \times \right. q_{T}\left. =\right. T$; (ii) $%
\mathbb{P}\left( Z_{i}^{\ast }\neq Z_{i}\right) \leq \beta (q_{T})$ for all $%
i\left. \in \right. \{1,\ldots ,k_{T}\}$; and (iii) $\{Z_{1}^{\ast
},Z_{3}^{\ast },\ldots \}$ are independent and $\{Z_{2}^{\ast },Z_{4}^{\ast
},\ldots \}$ are independent (cf.\ Lemma 2.1 in \cite{Berbee87} and
Proposition 2 in \cite{Doukan95}). Suppose $k_{T}$ is an even integer for
simplicity and define $U_{i}^{\ast }$ as i.i.d.\ standard uniform random
variable. Then these properties yield that 
\begin{align*}
& \text{ \ \ \ }\mathbb{P}\left( \min_{t\in \{1,\ldots
,T\}}\{D_{t}\}>\varepsilon T^{-\eta }\right) \\
& =\mathbb{P}\left( \min_{t\in \{1,\ldots ,T\}}\{D_{t}\}>\varepsilon
T^{-\eta },\{D_{t}\}_{t=1}^{T}=\{D_{t}^{\ast }\}_{t=1}^{T}\right) \\
& \text{ \ \ \ \ }+\mathbb{P}\left( \min_{t\in \{1,\ldots
,T\}}\{D_{t}\}>\varepsilon T^{-\eta },\{D_{t}\}_{t=1}^{T}\neq \{D_{t}^{\ast
}\}_{t=1}^{T}\right) \\
& \left. \leq _{(1)}\right. \mathbb{P}\left( \min_{t\in
\{2q_{T},4q_{T}\ldots ,k_{T}q_{T}\}}\{D_{t}^{\ast }\}>\varepsilon T^{-\eta
}\right) +\mathbb{P}\left( \{D_{t}\}_{t=1}^{T}\neq \{D_{t}^{\ast
}\}_{t=1}^{T}\right) \\
& \left. \leq _{(2)}\right. \mathbb{P}\left( \min_{i\in \{1,2,\ldots
,k_{T}/2\}}\{U_{i}^{\ast }\}>F_{D}\left( \varepsilon T^{-\eta }\right)
\right) +\mathbb{P}\left( \{D_{t}\}_{t=1}^{T}\neq \{D_{t}^{\ast
}\}_{t=1}^{T}\right) \\
& \left. \leq _{(3)}\right. (1-CT^{-\eta })^{k_{T}/2}+k_{T}\beta \left(
q_{T}\right) ,
\end{align*}%
where inequality(1) follows by considering the first elements in all even
blocks, which are independent by property(iii) above, inequality(2) follows
from the CDF transformation, and inequality(3) follows from the CDF of the
standard uniform distribution and properties (ii) and (iii) above. }

{\small Choosing $k_{T}$ as the largest even integer no larger than $2T^{1/3}
$ and using Condition 1.1 again yield that%
\begin{eqnarray*}
&&\sum_{T=1}^{\infty }\mathbb{P}\left( \min_{t\in \{1,\ldots
,T\}}\{D_{t}\}>\varepsilon T^{-\eta }\right) \\
&\leq &\sum_{T=1}^{\infty }(1-cT^{-\eta })^{T^{1/3}}+\sum_{T=1}^{\infty
}T^{1/3}O\left( T^{-4/3-2\varepsilon }\right) \\
&<&\infty \text{ for any }\eta \in (0,1/3).
\end{eqnarray*}%
Then $T^{\eta }\left\vert \left\vert X_{\left( x_{0}\right) }\left. -\right.
x_{0}\right\vert \right\vert \left. =\right. o_{a.s.}(1)$ is implied by
Borel Cantelli Lemma. The convergence of $\sum_{T=1}^{\infty }(1-cT^{-\eta
})^{T^{1/3}}$ is checked by the ratio test that $\lim_{T\rightarrow \infty
}(1\left. -\right. c\left( T\left. +\right. 1\right) ^{-\eta })^{\left(
T+1\right) ^{1/3}}/(1\left. -\right. cT^{-\eta })^{T^{1/3}}\left. <\right. 1$%
. Thus, (\ref{uniform o1}) holds with any $\eta \in (0,1/3)$. }

{\small Now we prove (\ref{mse o1}). Perform the same coupling argument as
above and consider the minimum value within each block $Z_{i}$ (and $%
Z_{i}^{\ast }$), denoted $\min \{Z_{i}\}$ (and $\min \{Z_{i}^{\ast }\}$).
Let $E_{T}$ denote the event that $\{D_{t}\}_{t=1}^{T}=\{D_{t}^{\ast
}\}_{t=1}^{T}$. The above three properties and (\ref{uniform o1}) yield that
for some constant $C>0$, 
\begin{align*}
& \text{ \ \ \ }\mathbb{E}[\left\vert \left\vert X_{\left( x_{0}\right)
}-x_{0}\right\vert \right\vert ] \\
& =\mathbb{E}\left[ \min_{t\in \{1,\ldots ,T\}}\{D_{t}\}\mathbf{1[}E_{T}%
\mathbf{]}\right] +\mathbb{E}\left[ \min_{t\in \{1,\ldots ,T\}}\{D_{t}\}%
\mathbf{1[}E_{T}^{c}\mathbf{]}\right] \\
& \left. \leq _{(1)}\right. \mathbb{E}\left[ \min_{i\in \{2,4,\ldots
,k_{T}\}}\{\min \{Z_{i}\}\}\mathbf{1[}E_{T}\mathbf{]}\right] +CT^{-\eta }%
\mathbb{E}\left[ \mathbf{1[}E_{T}^{c}\mathbf{]}\right] \\
& \left. \leq _{(2)}\right. \mathbb{E}\left[ \min_{i\in \{2,4,\ldots
,k_{T}\}}\{\min \{Z_{i}^{\ast }\}\}\right] +CT^{-\eta }k_{T}\beta \left(
q_{T}\right) \\
& \left. \leq _{(3)}\right. \mathbb{E}\left[ \min_{i\in \{2,4,\ldots
,k_{T}\}}\{D_{i\times q_{T}}^{\ast }\}\right] +CT^{-\eta }k_{T}\beta \left(
q_{T}\right) ,
\end{align*}%
where inequality(1) follows from considering even blocks only and (\ref%
{uniform o1}), inequality(2) follows from property(ii) above, and
inequality(3) follows from the fact that $\min \{Z_{i}^{\ast }\}\leq
D_{i\times q_{T}}^{\ast }$ (the minimum value within the block $Z_{i}^{\ast
} $ is less than or equal to the last element in that block). }

{\small The second term in the last step above is $o(T^{-1/2})$ by setting $%
q_{T}=k_{T}$ equal to the largest even integer no larger than $T^{1/2}$.
Regarding the first item above, notice that $\min_{i\in \{1,\ldots
,k_{T}\}}\{D_{i\times q_{T}}^{\ast }\}$ is the sample minimum of $k_{T}/2$
random samples from some CDF $F_{D}\left( \cdot \right) $, which has the
bounded lower end-point 0. Condition 1.1 implies that $F_{D}\left( \cdot
\right) $ is continuously differentiable and monotonically increasing in a
neighborhood of zero. Then we have 
\begin{align*}
\mathbb{E}\left[ \min_{i\in \{2,4,\ldots ,k_{T}\}}\{D_{i\times q_{T}}^{\ast
}\}\right] & =\mathbb{E}\left[ F_{D}^{-1}\left( \min_{i\in \{2,4,\ldots
,k_{T}\}}\{U_{i}^{\ast }\}\right) \right] \\
& \left. =_{(1)}\right. \mathbb{E}\left[ \left( 1/f_{D}\left( F_{D}^{-1}(%
\dot{u})\right) \right) \min_{i\in \{2,4,\ldots ,k_{T}\}}\{U_{i}^{\ast }\}%
\right] \\
& \left. \leq _{(2)}\right. C\mathbb{E}\left[ \min_{i\in \{2,4,\ldots
,k_{T}\}}\{U_{i}^{\ast }\}\right] \\
& \left. =_{(3)}\right. O(k_{T}^{-1}),
\end{align*}%
where equality(1) follows from mean value expansion with some $\dot{u}$
between $0$ and $\min_{i\in \{2,4,\ldots ,k_{T}\}}\{U_{i}^{\ast }\}$,
inequality(2) follows from the fact that $f_{D}\left( \cdot \right) $ is
uniformly bounded away from 0 in a neighborhood of zero, which is implied by
Condition 1.1 again, and equality(3) follows from Theorem 5.3.1 in \cite%
{deHaan07} since $U_{i}^{\ast }$ is i.i.d.\ standard uniform distribution
with the tail index $-1$. So (\ref{mse o1}) is established by setting $k_{T}$
equal to the largest even integer no larger than $T^{1/2}$ again. $%
\blacksquare $ \newline
}

{\small Now using Lemma \ref{lemma NN}, we prove Theorem \ref{thm main}. }

\paragraph{{ Proof of Theorem \protect\ref{thm main}}}

{\small Since the proof is long, we decompose it into three steps: (i) we
first establish the convergence in distribution of $Y_{(1),[x_{0}]}$ to $%
V_{1}$; (ii) we next generalize it to the whole vector $\mathbf{Y}$; and
(iii) we finally construct the test in the limiting problem as a function of 
$\mathbf{V}$ so that the uniform coverage is established by construction. \bigskip\newline
\textbf{Step 1.} We claim that, under Conditions 1.1-1.4, there exist
sequences of constants $a_{n}\left. >\right. 0$ and $b_{n}$ depending on $%
x_{0}$ such that 
\begin{equation}
\frac{Y_{(1),[x_{0}]}-b_{n}}{a_{n}}\overset{d}{\rightarrow }V_{1},
\label{evt_1}
\end{equation}
where $V_{1}$ is EV distributed with (\ref{def_G}) and $\xi =\xi (x_{0})$. }

{\small By Corollary 1.2.4 and Remark 1.2.7 in \cite{deHaan07}, the
constants $a_{n}$ and $b_{n}$ can be chosen as follows. If $\xi
(x_{0})\left. >\right. 0$, we choose $a_{n}(\xi (x_{0}))\left. =\right.
Q_{Y|X=x_{0}}(1\left. -\right. 1/n)$ and $b_{n}(\xi (x_{0}))\left. =\right. 0
$. If $\xi (x_{0})\left. =\right. 0$, we choose $a_{n}(\xi (x_{0}))\left.
=\right. 1/(nf_{Y|X=x_{0}}(b_{n}(x_{0})))$ and $b_{n}(\xi (x_{0}))\left.
=\right. Q_{Y|X=x_{0}}(1\left. -\right. 1/n)$. If }$\xi (x_{0})\left.
<\right. 0${\small , we choose }$a_{n}(\xi (x_{0}))\left. =\right. -\xi
\left( x_{0}\right) (y_{0}-Q_{Y|X=x_{0}}(1\left. -\right. 1/n))\left.
>\right. 0${\small \ and }$b_{n}(\xi (x_{0}))\left. =\right.
Q_{Y|X=x_{0}}(1\left. -\right. 1/n)${\small \ (Lemma 1.2.9 in \cite{deHaan07}%
), where recall that }$y_{0}${\small \ denotes the right end-point of }$%
F_{Y|X=x_{0}}${\small .\ By construction, these constants satisfy $1\left.
-\right. F_{Y|X=x_{0}}(a_{n}(\xi (x_{0}))y\left. +\right. b_{n}(\xi
(x_{0})))\left. =\right. O(n^{-1})$ for any fixed $y\left. >\right. 0$ in
both cases (cf.\ Chapter 1.1.2 in \cite{deHaan07}). }

{\small Let us suppress $\xi (x_{0})$ in the notations of $a_{n}\left( \cdot
\right) $ and $b_{n}\left( \cdot \right) $. By strict stationarity across $t$
(Condition 1.1), 
\begin{eqnarray}
\mathbb{P}\left( Y_{i,[x_{0}]}\leq v\right) &=&\mathbb{E}_{X_{i,\left(x_{0}%
\right) }}\left[ \mathbb{P}\left( Y_{i,[x_{0}]}\leq
v|X_{i,\left(x_{0}\right) }\right) \right]  \notag \\
&=&\mathbb{E}_{X_{i,\left( x_{0}\right) }}\left[ F_{Y|X=X_{i,(x_{0})}}%
\left(v\right) \right]  \label{key}
\end{eqnarray}
holds for any generic argument $v$. Thus, we have 
\begin{eqnarray*}
&&\mathbb{P}\left( Y_{(1),[x_{0}]}\leq a_{n}y+b_{n}\right) \\
&=&F_{Y_{i,[x_{0}]}}^{n}(a_{n}y+b_{n})\text{ (by i.i.d.\thinspace across }i%
\text{)} \\
&=&F_{Y|X=x_{0}}^{n}\left( a_{n}y+b_{n}\right) \left( \frac{\mathbb{P}%
\left(Y_{i,[x_{0}]}\leq a_{n}y+b_{n}\right) }{F_{Y|X=x_{0}}%
\left(a_{n}y+b_{n}\right) }\right) ^{n}\text{ } \\
&=&F_{Y|X=x_{0}}^{n}\left( a_{n}y+b_{n}\right) \left( \frac{\mathbb{E}%
_{X_{i,(x_{0})}}\left[ F_{Y|X=X_{i,(x_{0})}}\left( a_{n}y+b_{n}\right)\right]
}{F_{Y|X=x_{0}}\left( a_{n}y+b_{n}\right) }\right) ^{n}\text{ (by (\ref{key}%
))} \\
&=&F_{Y|X=x_{0}}^{n}\left( a_{n}y+b_{n}\right) \left( 1+\frac{\mathbb{E}%
_{X_{i,(x_{0})}}\left[ F_{Y|X=X_{i,(x_{0})}}\left( a_{n}y+b_{n}\right) %
\right] -F_{Y|X=x_{0}}\left( a_{n}y+b_{n}\right) }{F_{Y|X=x_{0}}%
\left(a_{n}y+b_{n}\right) }\right) ^{n} \\
&\equiv &A_{n}\left( y\right) \left( 1+\frac{B_{n,T}\left( y\right) }{%
F_{Y|X=x_{0}}\left( a_{n}y+b_{n}\right) }\right) ^{n}.
\end{eqnarray*}%
By the EV theory and Condition 1.2, $A_{n}\left( y\right)\rightarrow G_{\xi
}\left( y\right) $ as $n\rightarrow \infty $. Regarding $B_{n,T}\left(
y\right) $, we derive that, for some $\dot{x}_{i}$ between $X_{i,\left(
x_{0}\right) }$ and $x_{0}$ for each $i$, some open ball $B_{\eta
_{T}}(x_{0})$ centered at $x_{0}$ with radius $\eta _{T}\left. =\right.
O\left( T^{-\eta }\right) $, and some constant $0\left. <\right. C\left.
<\right. \infty $, 
\begin{align*}
\text{ \ \ \ }\left\vert B_{n,T}\left( y\right) \right\vert & \left.
=_{(1)}\right. \mathbb{E}\left[ \frac{\partial }{\partial x}F_{Y|X=x}\left(
a_{n}y+b_{n}\right) |_{x=\dot{x}_{i}}\left( X_{i,\left(x_{0}\right)
}-x_{0}\right) \right] \\
& \left. \leq _{(2)}\right. CT^{-\eta }\sup_{x\in B_{T^{-\eta
}}\left(x_{0}\right) }\left\vert \left\vert \frac{\partial }{\partial x}%
F_{Y|X=x}\left( a_{n}y+b_{n}\right) \right\vert \right\vert \\
& \left. \leq _{(3)}\right. CT^{-\eta }n^{-1}\sup_{x\in B_{T^{-\eta
}}\left(x_{0}\right) }\left\vert \left\vert \frac{\frac{\partial }{\partial x%
}F_{Y|X=x}\left( a_{n}y+b_{n}\right) }{1-F_{Y|X=x_{0}}(a_{n}y+b_{n})}%
\right\vert \right\vert \\
& \left. =_{(4)}\right. o(n^{-1})
\end{align*}
holds, where equality (1) is by the mean value expansion; inequality (2)
follows from that $X_{i,\left( x_{0}\right) }\in B_{\eta _{T}}(x_{0})$ holds
almost surely (Lemma \ref{lemma NN}); inequality (3) is due to $1\left.
-\right. F_{Y|X=x_{0}}(a_{n}y\left. +\right. b_{n})\left. =\right. O(1/n)$;
and equality (4) is given by Conditions 1.3-1.4. Hence given $%
a_{n}y+b_{n}\rightarrow y_{0}$ and using Lemma 8.4.1 in \cite{Arnold92}, we
have 
\begin{eqnarray*}
\left( 1+\frac{B_{n,T}\left( y\right) }{F_{Y|X=x_{0}}\left(a_{n}y+b_{n}%
\right) }\right) ^{n} &\leq &\left( 1+\frac{o\left(n^{-1}\right) }{%
F_{Y|X=x_{0}}\left( a_{n}y+b_{n}\right) }\right) ^{n} \\
&\rightarrow &1.
\end{eqnarray*}
The proof for the case of $k=1$ is then complete by the continuous mapping
theorem. \bigskip\newline
\textbf{Step 2.} We next claim that (\ref{evt_1}) can be generalized to the
cases of $k>1$ in the sense that, for the same $a_{n}$ and $b_{n}$ as in (%
\ref{evt_1}), the convergence (\ref{evt_k}) holds, that is, 
\begin{equation*}
\frac{\mathbf{Y-}b_{n}}{a_{n}}\overset{d}{\rightarrow }\mathbf{V}.
\end{equation*}
}

{\small To this end, consider $y_{1}>y_{2}>\cdots >y_{k}$. Theorem 8.4.2 in 
\cite{Arnold92} gives that 
\begin{eqnarray*}
&&\mathbb{P}\left( Y_{(1),[x_{0}]}\leq
a_{n}y_{1}+b_{n},...,Y_{(k),[x_{0}]}\leq a_{n}y_{k}+b_{n}\right) \\
&=&F_{Y_{i,[x_{0}]}}^{n-k}(a_{n}y_{k}+b_{n})\prod_{r=1}^{k}\left(
n-r+1\right) a_{n}f_{Y_{i},\left[ x_{0}\right] }\left(
a_{n}y_{r}+b_{n}\right) \text{ (by i.i.d. across }i\text{)} \\
&=&\left[ F_{Y|X=x_{0}}^{n-k}\left( a_{n}y_{k}+b_{n}\right)
\prod_{r=1}^{k}\left( n-r+1\right) a_{n}f_{Y|X=x_{0}}\left(
a_{n}y_{r}+b_{n}\right) \right] \times \\
&&\left[ \left( \frac{\mathbb{P}\left( Y_{i,[x_{0}]}\leq
a_{n}y_{k}+b_{n}\right) }{F_{Y|X=x_{0}}\left( a_{n}y_{k}+b_{n}\right) }%
\right) ^{n-k}\prod_{r=1}^{k}\frac{f_{Y_{i,\left[ x_{0}\right]
}}\left(a_{n}y_{r}+b_{n}\right) }{f_{Y|X=x_{0}}\left(
a_{n}y_{r}+b_{n}\right) }\right] \\
&\equiv &\tilde{A}_{n}\times \tilde{B}_{nT}.
\end{eqnarray*}
The convergence $\tilde{A}_{n}\rightarrow G_{\xi }\left(
y_{k}\right)\prod_{r=1}^{k}\{g_{\xi }\left( y_{r}\right) /G_{\xi }\left(
y_{k}\right) \} $ is established by Theorem 8.4.2 in \cite{Arnold92}. It now
remains to show $\tilde{B}_{nT}\rightarrow 1$. First, $(\mathbb{P}\left(
Y_{i,[x_{0}]}\leq a_{n}y_{k}+b_{n}\right) /F_{Y|X=x_{0}}\left(
a_{n}y_{k}+b_{n}\right) )^{n-k}\rightarrow 1$ is shown by the same argument
as above in the $k=1$ case. Second, for any $v$, we have 
\begin{eqnarray*}
\frac{f_{Y_{i,\left[ x_{0}\right] }}\left( v\right) }{f_{Y|X=x_{0}}\left(
v\right) } &=&\frac{\frac{\partial \mathbb{P}\left( Y_{i,[x_{0}]}\leq
v\right) }{\partial v}}{f_{Y|X=x_{0}}\left( v\right) } \\
&=&\frac{\frac{\partial }{\partial v}\mathbb{E}_{X_{i,(x_{0})}}\left[
F_{Y|X=X_{i,(x_{0})}}\left( v\right) \right] }{f_{Y|X=x_{0}}\left( v\right) }%
\text{ (by (\ref{key}))} \\
&=&\frac{\frac{\partial }{\partial v}\int F_{Y|X=x}\left( v\right)
f_{X_{i},\left( x_{0}\right) }\left( x\right) dx}{f_{Y|X=x_{0}}\left(
v\right) } \\
&=&\frac{\int \frac{\partial }{\partial v}F_{Y|X=x}\left( v\right)
f_{X_{i},\left( x_{0}\right) }\left( x\right) dx}{f_{Y|X=x_{0}}\left(
v\right) }\text{ \ (by Leibniz's rule)} \\
&=&\frac{\mathbb{E}_{X_{i,(x_{0})}}\left[ f_{Y|X=X_{i,(x_{0})}}\left(
v\right) \right] }{f_{Y|X=x_{0}}\left( v\right) },\text{ }
\end{eqnarray*}
where the application of Leibniz's rule is permitted under the assumption
(Condition 1.3) that $f_{Y|X=x}\left( v\right) $ is uniformly continuous in $%
x$ and $v$. Then similarly to the argument of bounding $B_{n,T}$ above, we
use the mean value expansion under Condition 1.3, Lemma \ref{lemma NN}, and
Conditions 1.3-1.4 to derive that for any $r\in \{1,...,k\}$ and some
constant $0\left. <\right. C\left.<\right. \infty $, 
\begin{eqnarray*}
&&\left\vert \frac{f_{Y_{i,\left[ x_{0}\right] }}\left(
a_{n}y_{r}+b_{n}\right) }{f_{Y|X=x_{0}}\left( a_{n}y_{r}+b_{n}\right) }%
-1\right\vert \\
&=&\left\vert \frac{\mathbb{E}_{X_{i,(x_{0})}}\left[ f_{Y|X=X_{i,(x_{0})}}%
\left( v\right) -f_{Y|X=x_{0}}\left( a_{n}y_{r}+b_{n}\right) \right] }{%
f_{Y|X=x_{0}}\left( a_{n}y_{r}+b_{n}\right) }\right\vert \\
&\leq &\sup_{x\in B_{\eta _{T}}(x_{0})}\left\vert \left\vert \frac{\partial
f_{Y|X=x}\left( a_{n}y_{r}+b_{n}\right) /\partial x}{f_{Y|X=x_{0}}\left(
a_{n}y_{r}+b_{n}\right) }\right\vert \right\vert \mathbb{E}\left[ \left\vert
\left\vert X_{i,(x_{0})}-x_{0}\right\vert \right\vert \right] \\
&\leq &o(1)\times O\left( T^{-\eta }\right) \\
&=&o(1).
\end{eqnarray*}
 \ \ \newline
\textbf{Step 3.} Note that $\lim_{n\rightarrow \infty
}F_{Y|X=x_{0}}(a_{n}v+b_{n}) = G_{\xi }(v)$, (\ref{evt_k}), and the
continuous mapping theorem yield that 
\begin{equation*}
\left(\begin{array}{c}
\frac{Q_{Y|X=x_{0}}(1-h/n)-Y_{(k),[x_{0}]}}{Y_{(1),[x_{0}]}-Y_{(k),[x_{0}]}}
\\
\frac{\mathbf{Y}-Y_{(k),[x_{0}]}}{Y_{(1),[x_{0}]}-Y_{(k),[x_{0}]}}
\end{array}\right) 
\overset{d}{\rightarrow }
\left(\begin{array}{c}
V^{q}
\\
\mathbf{V}^{\ast }
\end{array}\right).
\end{equation*}
By another application of the continuous mapping theorem, any equivariant $S$
that satisfies the asymptotic size constraint $\mathbb{P}(V^{q}\in S(\mathbf{%
V}^{\ast }))\left. \geq \right. 1\left. -\right. \alpha $ for every $\xi \in
\Xi $ also satisfies $\lim \inf_{n\rightarrow \infty ,T\rightarrow \infty }%
\mathbb{P}(Q_{Y|X=x_{0}}(1-h/n)\in S(\mathbf{Y}))\left. \geq \right.
1\left.-\right. \alpha $. Thus, it suffices to determine $S\left( \cdot
\right) $ in the limiting problem where the observation is $\mathbf{V}^{\ast
}$. Given the solution (\ref{S_Lambda}), it further suffices to determine a
suitable Lagrangian weight $\Lambda $. We accomplish this by construction. }

{\small Consider $\Lambda =c\tilde{\Lambda}$, where $\tilde{\Lambda}$ is
some probability distribution function with support on $\Xi $ and $c$ some
positive constant to be determined. Note that the density $f_{V^{q},\mathbf{V%
}^{\ast }}(y,\mathbf{v}^{\ast };\xi )$ is continuously differentiable in all
three arguments, and hence $\mathbb{P}_{\xi }\left( V^{q}\in S_{c\tilde{%
\Lambda}}(\mathbf{V}^{\ast })\right) $ as a function of $\xi $ and $c$ is
continuous in both arguments. Denote by $S$ in (\ref{S_Lambda}) as $%
S_{\Lambda }$ to indicate that the confidence interval depends on the choice
of $\Lambda $. Then, given any $\tilde{\Lambda}$ and $c$, $\inf_{\xi \in \Xi
}\mathbb{P}_{\xi }\left( V^{q}\in S_{c\tilde{\Lambda}}(\mathbf{V}^{\ast
})\right) $ is obtainable, say at $\xi ^{\ast }$, since $\Xi $ is compact.
Furthermore, for every $\xi \in \Xi $, $\mathbb{P}_{\xi }\left( V^{q}\in S_{c%
\tilde{\Lambda}}(\mathbf{V}^{\ast })\right) $ as a function of $c$ is
increasing. We can choose $c=c^{\ast }$ such that $\mathbb{P}_{\xi ^{\ast
}}\left(V^{q}\in S_{c^{\ast }\tilde{\Lambda}}(\mathbf{V}^{\ast })\right)
=1-\alpha $. This is always feasible since $\inf_{\xi \in \Xi }\mathbb{P}%
_{\xi }\left( V^{q}\in S_{c^{\ast }\tilde{\Lambda}}(\mathbf{V}^{\ast
})\right) \rightarrow 1$ as $c^{\ast }\rightarrow \infty $ and $\sup_{\xi
\in \Xi }\mathbb{P}_{\xi }\left( V^{q}\in S_{c^{\ast }\tilde{\Lambda}}(%
\mathbf{V}^{\ast })\right) \rightarrow 0$ as $c^{\ast }\rightarrow 0$. The
proof is then complete since $\tilde{\Lambda}$ can be arbitrary. $%
\blacksquare $ }

\begin{remark}
{\small Since $\tilde{\Lambda}$ in the last part of the above proof can be
arbitrary in theory, we provide an empirical guide for determining a nearly
optimal $\tilde{\Lambda}$ in Section \ref{sec computation}. }
\end{remark}

\paragraph{{ Proof of Corollary \protect\ref{col ab}}}

{\small By Corollary 1.2.4 and Remark 1.2.7 in \cite{deHaan07}, the
constants $a_{n}$ and $b_{n}$ can be chosen as follows. We present the case
for $\alpha $ only, and the choice for $\beta $ follows identically. If $\xi
_{\alpha }\left. >\right. 0$, we choose $a_{n}(\xi _{\alpha })\left.
=\right. Q_{\alpha }(1\left. -\right. 1/n)$ and $b_{n}(\xi _{\alpha })\left.
=\right. 0$, where recall that $Q_{\alpha }(\cdot )$ denotes the quantile
function of $\alpha _{i}$. If $\xi _{\alpha }\left. =\right. 0$, we choose $%
a_{n}(\xi _{\alpha })\left. =\right. 1/(nf_{\alpha }(b_{n}(\xi _{\alpha })))$
and $b_{n}(\xi _{\alpha })\left. =\right. Q_{\alpha }(1\left. -\right. 1/n)$%
, where recall that $f_{\alpha }(\cdot )$ denotes the PDF of $\alpha _{i}$.
If }$\xi _{\alpha }<0${\small , we choose }$a_{n}\left( \xi _{\alpha
}\right) =-\xi _{\alpha }\left( Q_{\alpha }(1)-Q_{\alpha }(1\left. -\right.
1/n)\right) ${\small \ and }$b_{n}\left( \xi _{\alpha }\right) =Q_{\alpha
}(1\left. -\right. 1/n)$. {\small By construction, these constants satisfy
that $1\left. -\right. F_{\alpha }(a_{n}(\xi _{\alpha })y\left. +\right.
b_{n}(\xi _{\alpha }))\left. =\right. O(n^{-1})$ for any fixed $y\left.
>\right. 0$ in both cases (cf. Chapter 1.1.2 in \cite{deHaan07}). }

{\small We first establish the convergence of $\mathbf{A}$. By the EV
theory, Condition 2.1 ($\alpha _{i}$ is i.i.d.) and Condition 2.2 ($%
F_{\alpha }\left. \in \right. \mathcal{D}\left( G_{\xi _{\alpha }}\right) $)
imply 
\begin{equation}
\left( \frac{\alpha _{(1)}-b_{n}(\xi _{\alpha })}{a_{n}(\xi _{\alpha })},...,%
\frac{\alpha _{(k)}-b_{n}(\xi _{\alpha })}{a_{n}(\xi _{\alpha })}\right)
^{\intercal }\overset{d}{\rightarrow }\mathbf{V}\left( \xi _{\alpha }\right)
,  \label{EVT1}
\end{equation}
where $\mathbf{V}(\xi _{\alpha })$ is jointly EV distributed with tail index 
$\xi _{\alpha }$. }

{\small Let $I=(I_{1},\ldots ,I_{k})\in \{1,\ldots ,T\}^{k}$ be the $k$
random indices such that $\alpha _{\left( j\right) }=\alpha _{I_{j}}$, $%
j=1,\ldots ,k$, and let $\hat{I}$ be the corresponding indices such that $%
\hat{\alpha}_{(j)}=\hat{\alpha}_{\hat{I}_{j}}$. Then, the convergence of $%
\mathbf{A}$ follows from (\ref{EVT1}) once we establish $|\hat{\alpha}_{\hat{%
I}_{j}}-\alpha _{I_{j}}|=o_{p}(a_{n}\left( \xi _{\alpha }\right) )$ for $%
j=1,\ldots ,k$. We present the case of $k=1$, but the argument for a general 
$k$ is similar. Denote $\varepsilon _{i}\equiv \hat{\alpha}_{i}-\alpha _{i}$%
. }

{\small First, consider the case with $\xi _{\alpha }>0$. The part in
Condition 2.3 for $\xi _{\alpha }\left. >\right. 0$ yields that 
\begin{eqnarray*}
\sup_{i}\left\vert \varepsilon _{i}\right\vert &=&\sup_{i}\left\vert \bar{X}%
_{i}^{\intercal }\left( \beta _{i}-\hat{\beta}_{i}\right) +\bar{u}%
_{i}\right\vert \\
&\leq &\sup_{i}\left\vert \left\vert \bar{X}_{i}\right\vert \right\vert
\sup_{i}\left\vert \left\vert \beta _{i}-\hat{\beta}_{i}\right\vert
\right\vert +\sup_{i}\left\vert \bar{u}_{i}\right\vert \\
&=&o_{p}(1).
\end{eqnarray*}
Given this result, we have that, on one hand, $\hat{\alpha}_{\hat{I}} =
\max_{i}\{\alpha _{i} +\varepsilon _{i}\} \leq \alpha _{I} +
\sup_{i}\left\vert \varepsilon _{i}\right\vert = \alpha _{I} + o_{p}(1)$;
and, on the other hand, $\hat{\alpha}_{\hat{I}} = \max_{i}\{\alpha_{i}\left.
+\right. \varepsilon _{i}\} \geq \max_{i}\{\alpha_{i}+ \min_{i}\{\varepsilon
_{i}\}\} \geq \alpha_{I} + \min_{i}\{\varepsilon _{i}\} \geq \alpha_{I} -
\sup_{i}\left\vert \varepsilon _{i}\right\vert = \alpha _{I} - o_{p}(1)$.
Therefore, $\left\vert \hat{\alpha}_{\hat{I}}-\alpha _{I}\right\vert \leq
o_{p}(1)=o_{p}(a_{n}\left( \xi _{\alpha }\right) )$ since $a_{n}(\xi
_{\alpha })\rightarrow \infty $. }

{\small Second, consider the case with $\xi _{\alpha }=0$. Corollary 1.2.4
in \cite{deHaan07} implies that $a_{n}\left( \xi _{\alpha }\right)
=f_{\alpha }\left( Q_{\alpha }(1-1/n)\right) $. Thus, the part in Condition
2.3 for $\xi _{\alpha }\left. =\right. 0$ implies that 
\begin{eqnarray*}
\frac{1}{a_{n}\left( \xi _{\alpha }\right) }\sup_{i}\left\vert \varepsilon
_{i}\right\vert  &\leq &\frac{\sup_{i}\left\vert \left\vert \bar{X}%
_{i}\right\vert \right\vert \sup_{i}\left\vert \left\vert \beta _{i}-\hat{%
\beta}_{i}\right\vert \right\vert +\sup_{i}\left\vert \bar{u}_{i}\right\vert 
}{f_{\alpha }\left( Q_{\alpha }(1-1/n)\right) } \\
&=&o_{p}(1).
\end{eqnarray*}%
Now, the same argument as above yields that $\left\vert \hat{\alpha}_{\hat{I}%
}-\alpha _{I}\right\vert \left. \leq \right. O_{p}\left( \sup_{i}\left\vert
\varepsilon _{i}\right\vert \right) \left. =\right. o_{p}(a_{n}\left( \xi
_{\alpha }\right) )$. }

{\small Third, consider the case with }$\xi _{\alpha }<0${\small . The fact
that\ }$a_{n}(\xi _{\alpha })=O\left( n^{\xi _{\alpha }}\right) $\ implies
that 
\begin{eqnarray*}
\frac{1}{{\small a}_{n}\left( \xi _{\alpha }\right) }\sup_{i}\left\vert
\varepsilon _{i}\right\vert  &\leq &n^{-\xi _{\alpha }}\left(
\sup_{i}\left\vert \left\vert \bar{X}_{i}\right\vert \right\vert
\sup_{i}\left\vert \left\vert \beta _{i}-\hat{\beta}_{i}\right\vert
\right\vert +\sup_{i}\left\vert \bar{u}_{i}\right\vert \right)  \\
&=&O_{p}\left( n^{-\xi _{\alpha }}T^{-1/2}\right)  \\
&=&o_{p}(1)\text{,}
\end{eqnarray*}%
{\small where the two equalities follow from Condition 2.3. The rest of the
proof is identical to Step 3 in the proof of Theorem \ref{thm main}. }

{\small Finally, we establish the convergence of $\mathbf{B}$. Recall that
we focus on, without loss of generality, the first component of $\beta _{i}$%
, so that $(\beta _{(1)},...,\beta _{(k)})^{\intercal }$ denotes the largest 
$k$ elements in the first components of $\{\beta _{i}\}_{i=1}^{n}$.
Conditions 2.1 and 2.2 imply that 
\begin{equation*}
\left( \frac{\beta _{(1)}-b_{n}\left( \xi _{\beta }\right) }{a_{n}\left( \xi
_{\beta }\right) },...,\frac{\beta _{(k)}-b_{n}\left( \xi _{\beta }\right) }{%
a_{n}\left( \xi _{\beta }\right) }\right) ^{\intercal }\overset{d}{%
\rightarrow }\mathbf{V}\left( \xi _{\beta }\right) .
\end{equation*}
Condition 2.3 and a similar argument to that for $\mathbf{A}$ complete the
proof. $\blacksquare $ }

\subsection{{ Computational details\label{sec computation}}}

{\small We discuss the choice of $\Lambda $ following \cite{EMW15} and \cite%
{MuellerWang17}. Using the same notation as in the proof of Theorem \ref{thm
main}, consider $\Lambda =c\tilde{\Lambda}$, where $\tilde{\Lambda}$ is some
probability distribution function with support on $\Xi $. Suppose $\xi $ is
randomly drawn from $\tilde{\Lambda}$ and $c$ satisfies that $\int \mathbb{P}%
_{\xi }(V^{q}\in S_{c\tilde{\Lambda}}(\mathbf{V}^{\ast }))d \tilde\Lambda (\xi
)=1-\alpha $. Denote the $W$-weighted average length as $V_{\tilde{\Lambda}%
}=\int_{\Xi }\mathbb{E}_{\xi }[\kappa (\mathbf{V}^{\ast };\xi )\text{lgth}%
(S_{c\tilde{\Lambda}}(\mathbf{V}^{\ast }))]dW(\xi )$. Since the uniform
coverage for all $\xi \in \Xi $ implies the $\tilde{\Lambda}$-weighted
average coverage for any probability distribution $\tilde{\Lambda}$ and $S_{c%
\tilde{\Lambda}}$ minimizes the $W$-weighted average length by construction, 
$V_{\tilde{\Lambda}}$ essentially provides a lower bound for the $W$%
-weighted average length among all sets $S$ that satisfy the uniform
coverage. }

{\small Now suppose we obtain some $\tilde{\Lambda}^{\ast }$ on $\Xi $ and
the constant $c^{\ast }$ such that }

{\small 
\begin{equation}
\mathbb{P}_{\xi }(V^{q}\in S_{c^{\ast }\tilde{\Lambda}^{\ast }}(\mathbf{V}%
^{\ast }))\left. \geq \right. 1-\alpha \text{ for all }\xi \in \Xi \text{,}
\label{constraint1}
\end{equation}
and%
\begin{equation}
\int_{\Xi }\mathbb{E}_{\xi }[\kappa (\mathbf{V}^{\ast };\xi )\text{lgth}%
(S_{c^{\ast }\tilde{\Lambda}^{\ast }}(\mathbf{V}^{\ast }))]dW(\xi )\left.
\leq \right. (1+\varepsilon )V_{\tilde{\Lambda}^{\ast }},
\label{constraint2}
\end{equation}%
then the confidence interval $S_{c^{\ast }\tilde{\Lambda}^{\ast }}$ will
have a $W$-weighted average expected length no more than 100$\varepsilon \%$
longer than any other confidence interval of the same level. We set $%
\varepsilon =0.01$. }

{\small To identify a suitable choice of $\tilde{\Lambda}^{\ast }$, we can
discretize $\Xi $ into a grid $\Xi _{a}$ and determine $\tilde{\Lambda}$
accordingly as the point masses. Then we can simulate $N$ random draws of $%
\left( V^{q},\mathbf{V}^{\ast }\right) $ from $\xi \in \Xi _{a}$ and
estimate $\mathbb{P}_{\xi }(V^{q}\in S_{c^{\ast }\tilde{\Lambda}}(\mathbf{V}%
^{\ast }))$ by sample fractions. By iteratively increasing or decreasing the
point masses as a function of whether the estimated $\mathbb{P}_{\xi
}(V^{q}\in S_{c^{\ast }\tilde{\Lambda}}(\mathbf{V}^{\ast }))$ is larger or
smaller than the nominal level, we can always find a candidate $\tilde{%
\Lambda}^{\ast }$. Note that such $\tilde{\Lambda}^{\ast }$ always exists
since we allow $\mathbb{P}_{\xi }(V^{q}\in S_{c^{\ast }\tilde{\Lambda}}(%
\mathbf{V}^{\ast }))>1-\alpha $ for some $\xi $. We determine $c^{\ast }$ so
that (\ref{constraint2}) is satisfied. The continuity of $f_{V^{q},\mathbf{V}%
^{\ast }}(y,\mathbf{v}^{\ast };\xi )$ entails that $\mathbb{P}_{\xi
}(V^{q}\in S_{c^{\ast }\tilde{\Lambda}}(\mathbf{V}^{\ast }))$ as a function
of $\xi $ is also continuous. Therefore, (\ref{constraint1}) is guaranteed
as we consider $\left\vert \Xi _{a}\right\vert \rightarrow \infty $ and $%
N/\left\vert \Xi _{a}\right\vert \rightarrow \infty $, where $\left\vert \Xi
_{a}\right\vert $ denotes the cardinality of $\Xi _{a}$. }

{\small In our simulations, we consider $\Xi =[-1/2,1/2]$, $\Xi
_{a}=\{-1/2,-1/2+1/59,\ldots ,1/2\}$, and accordingly $\tilde{\Lambda}$ is
equal to 60 point masses on $\Xi _{d}$. We consider $\alpha =0.05$.
Following \cite{EMW15}, we determine these point masses by the following
steps. }

\begin{enumerate}
\item {\small Simulate $N=100,000$ i.i.d. random draws from some proposal
density with $\xi $ drawn uniformly from $\Xi _{p}=\{-1/2,-1/2+1/29,\ldots
,1/2\}$. }

\item {\small Start with $\tilde{\Lambda}_{(0)}=\{1/60,1/60,\ldots ,1/60\}$
and $c^{\ast }=1$. Calculate the (estimated) coverage probabilities $\mathbb{%
P}_{\xi _{j}}(V^{q}\in S_{c^{\ast }\tilde{\Lambda}(0)}(\mathbf{V}^{\ast }))$
for every $\xi _{j}\in \Xi _{a}$ using importance sampling. Denote them as $%
\mathbf{P=(}P_{1},...,P_{60})^{\prime }.$ }

\item {\small Update $\Lambda $ by setting $\Lambda _{(s+1)}=\Lambda
_{(s)}+\eta _{\Lambda }(\mathbf{P-}0.95)$ with some step-length constant $%
\eta _{\Lambda }>0$, so that the $j$-th point mass in $\Lambda $ is
increased/decreased if the coverage probability for $\xi _{j}$ is
larger/smaller than the nominal level. }

\item {\small Keep the integration for 500 times. Then the resulting $%
\Lambda _{(500)}$ is a valid candidate. }

\item {\small Numerically check if $S_{\Lambda _{(500)}}$ indeed controls
the coverage uniformly by simulating the coverage probabilities over the
fine enough grid $\Xi _{f}=\{-1/2,-1/2+1/199,\ldots ,1/2\}$. If not, go back
to step 2 with a finer $\Xi _{a}$. }
\end{enumerate}

{\small The expressions of $f_{V^{q},\mathbf{V}^{\ast }}$ and $\kappa (%
\mathbf{v}^{\ast };\xi )f_{\mathbf{V}^{\ast }}(\mathbf{v}^{\ast };\xi )$ are
as follows.%
\begin{eqnarray*}
f_{V^{q},\mathbf{V}^{\ast }}(y,\mathbf{v}^{\ast };\xi ) &=&\left\vert
y\right\vert ^{k}\int_{a_{1}(\xi )}^{b_{1}(\xi )}\left\vert q(\xi
,h)-s\right\vert ^{k-1} \\
&&\times \exp \left[ -(1+\xi s)^{-1/\xi }-(1+1/\xi )\right] \\
&&\times \sum_{i=1}^{k}\log \left( 1+\xi s+\mathbf{v}_{i}^{\ast }\xi \frac{%
q(\xi ,h)-s}{y}\right) ds,
\end{eqnarray*}%
where $a_{1}(\xi )$ and $b_{1}(\xi )$ are such that for all $s\in (a_{1}(\xi
),b_{1}(\xi ))$, $1+\xi s>0,1+\xi s+\xi (q(\xi ,h)-s)/y>0$. 
\begin{eqnarray*}
\kappa (\mathbf{v}^{\ast };\xi )f_{\mathbf{V}^{\ast }}(\mathbf{v}^{\ast
};\xi ) &=&\Gamma \left( k-\xi \right) \int_{0}^{b_{0}(\xi )}s^{k-1} \\
&&\times \exp \left[ -\left( 1+1/\xi \right) \sum_{i=1}^{k}\log \left( 1+\xi 
\mathbf{v}_{i}^{\ast }s\right) \right] ds,
\end{eqnarray*}%
where $\Gamma $ is the Gamma function, and $b_{0}(\xi )=-1/\xi $ for $\xi <0$
and $b(\xi )=\infty $ otherwise. }

\subsection{ Primitive conditions for Condition 1.3}\label{sec:primitive_conditions}

{\small In this appendix, we provide primitive conditions for Condition 1.3.
The following conditions are sufficient. Recall that $y_{0}$ denotes the
right end-point $\sup \{y,F_{Y|X=x_{0}}(y)\left. <\right. 1\}$. The notation
is simpler if we use the following notations: $\gamma (\cdot )\left.
=\right. 1/\xi (\cdot )$, $g_{i}$ denotes the the partial derivative of a
generic function $g(\cdot ,\cdot )$ w.r.t.\thinspace the $i$-th element, and 
$g_{ij}\ $the $i$,$j$-th cross derivative. }

\begin{description}
\item[Condition B] {\small $X_{it}$ has a compact support. $F_{Y|X=x}(y)$
satisfies one of the following three cases: (i) $\xi (x)>0$ and 
\begin{equation*}
1-F_{Y|X=x}(y)=c(x)y^{-\gamma (x)}(1+d(x)(y)^{-\tilde{\gamma}(x)}+r(x,y))
\end{equation*}%
where $c(\cdot )\left. >\right. 0$ and $d(\cdot )$ are uniformly bounded
between $0$ and $\infty $ and continuously differentiable with uniformly
bounded derivatives, $\gamma (\cdot )\left. >\right. 0$ and $\tilde{\gamma}%
(\cdot )\left. >\right. 0$ are continuously differentiable functions, and $%
r(x,y)$ is continuously differentiable with bounded derivatives w.r.t.\ both 
$x$ and $y,$ and satisfies for some $\delta >0$%
\begin{eqnarray*}
\underset{y\rightarrow y_{0}}{\lim \sup }\sup_{x\in B_{\delta }(x_{0})\cap
\{x:\xi (x)>0\}}\left\vert r(x,y)/y^{-\tilde{\gamma}(x)}\right\vert 
&\rightarrow &0, \\
\underset{y\rightarrow y_{0}}{\lim \sup }\sup_{x\in B_{\delta }(x_{0})\cap
\{x:\xi (x)>0\}}\left\vert r_{2}(x,y)/(y^{-\tilde{\gamma}(x)-1})\right\vert 
&\rightarrow &0, \\
\underset{y\rightarrow y_{0}}{\lim \sup }\sup_{x\in B_{\delta }(x_{0})\cap
\{x:\xi (x)>0\}}\left\vert r_{1}(x,y)/y^{-\tilde{\gamma}(x)}\right\vert 
&\rightarrow &0, \\
\underset{y\rightarrow y_{0}}{\lim \sup }\sup_{B_{\delta }(x_{0})\cap
\{x:\xi (x)>0\}}\left\vert r_{21}(x,y)/(y^{-\tilde{\gamma}(x)-1})\right\vert
&\rightarrow &0\text{;}
\end{eqnarray*}%
(ii) $\xi (x)=0$ and%
\begin{equation*}
f_{Y|X=x}(y)=c(x)y^{\tilde{c}(x)}\exp (-d(x)\tilde{d}(y))(1+r(x,y)),\text{ }
\end{equation*}%
where $c(\cdot )\left. >\right. 0$ and $d(\cdot )\left. >\right. 0$ are some
continuously differential functions that are uniformly bounded between $0$
and $\infty $, $\tilde{c}(\cdot )$ is continuously differentiable and
uniformly bounded by $-1$ and $\infty $, and $\tilde{d}(y)$ is continuously
differentiable and satisfies $C_{1}(\log y)^{2}\leq \tilde{d}(y)\leq
C_{2}y^{C_{3}}$ for some constants $0\leq C_{1},C_{2},C_{3}<\infty $. The
remainder $r(x,y)$ is uniformly bounded and continuously differentiable
w.r.t.\thinspace both arguments with bounded derivatives, and satisfies that
for some $\delta >0$ 
\begin{equation*}
\underset{y\rightarrow y_{0}}{\lim \sup }\sup_{x\in B_{\delta }(x_{0})\cap
\{x:\xi (x)=0\}}\left\vert \max
\{r_{1}(x,y),r_{2}(x,y),r_{21}(x,y)\}\right\vert \rightarrow 0.
\end{equation*}%
(iii) }$\xi (x)<0${\small , }%
\begin{equation*}
1-F_{Y|X=x}(y)=c(x)(y_{0}-y)^{-\gamma (x)}(1+d(x)(y_{0}-y)^{-\tilde{\gamma}%
(x)}+r(x,y))
\end{equation*}%
{\small where }$c(\cdot )\left. >\right. 0${\small \ and }$d(\cdot )${\small %
\ are uniformly bounded and continuously differentiable with uniformly
bounded derivatives, and }$\gamma (\cdot )\left. <\right. 0${\small \ and }$%
\tilde{\gamma}(\cdot )\left. <\right. 0${\small \ are continuously
differentiable functions, and }$r(x,y)${\small \ is continuously
differentiable with bounded derivatives w.r.t.\ both }$x${\small \ and }$y,$%
{\small \ and satisfies for some }$\delta >0$%
\begin{eqnarray*}
\underset{y\rightarrow y_{0}}{\lim \sup }\sup_{x\in B_{\delta }(x_{0})\cap
\{x:\xi (x)<0\}}\left\vert r(x,y)/(y_{0}-y)^{-\tilde{\gamma}(x)}\right\vert 
&\rightarrow &0, \\
\underset{y\rightarrow y_{0}}{\lim \sup }\sup_{x\in B_{\delta }(x_{0})\cap
\{x:\xi (x)<0\}}\left\vert r_{2}(x,y)/((y_{0}-y)^{-\tilde{\gamma}%
(x)-1})\right\vert  &\rightarrow &0, \\
\underset{y\rightarrow y_{0}}{\lim \sup }\sup_{x\in B_{\delta }(x_{0})\cap
\{x:\xi (x)<0\}}\left\vert r_{1}(x,y)/(y_{0}-y)^{-\tilde{\gamma}%
(x)})\right\vert  &\rightarrow &0, \\
\underset{y\rightarrow y_{0}}{\lim \sup }\sup_{x\in B_{\delta }(x_{0})\cap
\{x:\xi (x)<0\}}\left\vert r_{21}(x,y)/(y_{0}-y)^{-\tilde{\gamma}%
(x)-1})\right\vert  &\rightarrow &0\text{.}
\end{eqnarray*}
\end{description}

{\small Condition B assumes that the error of approximating the true CDF
with a generalized Pareto distribution consists of the leading terms $%
1+d(x)(y)^{-\tilde{\gamma}(x)}$ and $c(x)y^{\tilde{c}(x)}\exp (-d(x)\tilde{d}%
(y))$, respectively in the two cases with $\xi (x)>0$ and $\xi (x)=0$ and
the remainder $r(x,y)$. Most of it is essentially a conditional version of
the unconditional second order assumptions that are common in the EV
literature. In particular, Case (i) covers regularly varying tails, and are
imposed by \cite{Smith82} to study unconditional problems. See also \cite%
{Hall82} and \cite{Smith87}. Case (ii) covers slowly varying tails,
including Gaussian ($\tilde{c}(x)\left. =\right. 0$ and $\tilde{d}(y)\left.
=\right. y^{2}$), lognormal ($\tilde{c}(x)\left. =\right. -1$ and $\tilde{d}%
(y)\left. =\right. \left( \log y\right) ^{2}$), and the exponential family ($%
\tilde{c}(x)\left. =\right. 0$ and $\tilde{d}(y)\left. =\right. y$). See,
for example, Chapter B in \cite{deHaan07}. Case (iii) covers the thin tail
case where the conditional distribution has a bounded right end-point. For
example, the standard uniform distribution on }$[0,1]${\small \ is covered
with }$c(x)=$ $y_{0}=$ $-\gamma (x)=1$ and $d(x)=0$. {\small Compared with
the unconditional EV literature, we require a stronger version that the
derivatives of $r(x,y)$ are uniformly bounded. This is to guarantee that the
tail of $f_{Y|X=x_{0}}$ is also uniformly bounded. The compact support of $X$
is imposed to simplify the proof (cf.\thinspace \cite{WangLi13}). The
following lemma establishes Condition 1.3 using Conditions 1.4 and B. Its
proof is collected at the very end of this article. }

\begin{lemma}
{\small \label{lemma: pre}If Condition 1.4 and Condition B hold, then
Condition 1.3 holds, i.e., for $u_{n}\left. =\right. a_{n}y\left. +\right.
b_{n}$ with any fixed $y\left. >\right. 0$, as $n\left. \rightarrow \right.
\infty $ and $T\left. \rightarrow \right. \infty $ }

{\small (a) $\lim_{u_{n}\rightarrow y_{0}}\sup_{x\in B_{\eta
_{T}}(x_{0})}T^{-\eta }\left\vert \left\vert \frac{\partial F_{Y|X=x}\left(
u_{n}\right) /\partial x}{1-F_{Y|X=x_{0}}(u_{n})}\right\vert \right\vert
\left. =\right. 0$, }

{\small (b) $\lim_{u_{n}\rightarrow y_{0}}\sup_{x\in B_{\eta
_{T}}(x_{0})}T^{-\eta }\left\vert \left\vert \frac{\partial f_{Y|X=x}\left(
u_{n}\right) /\partial x}{f_{Y|X=x}\left( u_{n}\right) }\right\vert
\right\vert \left. =\right. 0$. }
\end{lemma}

{\small To give a better sense of Condition B, we now show that it is
satisfied by the three examples introduced in Section \ref{sec general}. }

{\small First consider the joint normal distribution. Condition B.(ii) is
satisfied by setting $c(x)=\sqrt{2\pi (1-\rho ^{2})}$,$\ d(x)=1$, $\tilde{d}%
(y)=y^{2}/(2(1-\rho ^{2}))$, and $r(x,y)=\exp (2\rho x/y+\rho
^{2}x^{2}/y^{2})-1$. Second, for the conditional Student-t distribution, 
\cite{Ding16} derives that the conditional PDF\ of $Y$ given $X=x$ is 
\begin{equation*}
f_{Y|X=x}(y)=\frac{C}{\sigma (x)}\left( 1+\frac{(y-\rho x)^{2}}{\left(
v+1\right) \sigma (x)^{2}}\right) ^{-\frac{v+2}{2}}
\end{equation*}%
for some constant $C$ depending on $v$ only and $\sigma (x)\left. =\right. 
\sqrt{(1-\rho ^{2})(v+x^{2})/(v+1)}$. Then Condition B.(i) holds with $%
\gamma (x)=1/(v+1)$, $c(x)\left. \propto \right. \sigma (x)^{v+1},d(x)\left.
\propto \right. \rho x,\tilde{\gamma}(x)\left. =\right. 1$, and $%
r(x,y)\left. =\right. O(y^{-2})$ for any $x\in 
\mathbb{R}
$. Finally, for the conditional Pareto distribution, Taylor expansion yields 
\begin{eqnarray*}
1-F_{Y|X=x}(y) &=&y^{-1/x}(1+1/y)^{-1/x} \\
&=&y^{-1/x}(1-\frac{1}{xy}+O(\frac{1}{y^{2}})).
\end{eqnarray*}%
Thus Condition B.(i) holds with $c(x)=1,\gamma (x)=1/x,d(x)=-1/x,\tilde{%
\gamma}(x)=1,$ and $r(x,y)=O(y^{-2})$ for $x$ bounded below from 0. }

\paragraph{{ Proof of Lemma \protect\ref{lemma: pre}}}

{\small The proof is different for $\xi (x_{0})\left. >\right. $ $=\left.
<\right. 0$.. We first consider the positive $\xi (x_{0})$ case. Recall that 
$B_{\eta _{T}}(x_{0})$ denotes an open ball centered at $x_{0}$ with radius $%
\eta _{T}=T^{-\eta }$, where $\eta $ is determined in Lemma \ref{lemma NN}.
For (a), $\inf_{x\in B_{\eta _{T}}(x_{0})}\xi (x)>0$ if $T$ is large enough.
This is feasible given the continuity of $\xi (\cdot )$. Then by the chain
rule and the condition (Condition B.(i)) that 
\begin{equation}
1-F_{Y|X=x}(y)=c(x)y^{-\gamma (x)}(1+d(x)(y)^{-\tilde{\gamma}(x)}+r(x,y)),
\label{cond Bi1}
\end{equation}%
we have%
\begin{eqnarray*}
&&\frac{\partial F_{Y|X=x}\left( y\right) /\partial x}{1-F_{Y|X=x}\left(
y\right) } \\
&=&\frac{c_{1}(x)}{c(x)}-\gamma _{1}(x)\log y+\frac{d_{1}(x)y^{-\tilde{\gamma%
}(x)}}{1+d(x)(y)^{-\tilde{\gamma}(x)}+r(x,y)} \\
&&-\frac{d(x)y^{-\tilde{\gamma}(x)}\tilde{\gamma}_{1}(x)\log y}{1+d(x)(y)^{-%
\tilde{\gamma}(x)}+r(x,y)}+\frac{r_{1}(x,y)}{1+d(x)(y)^{-\tilde{\gamma}%
(x)}+r(x,y)}.
\end{eqnarray*}%
Recall that 
\begin{eqnarray}
u_{n} &=&a_{n}y\left. +\right. b_{n}  \notag \\
&=&O(Q_{Y|X=x_{0}}(1\left. -\right. 1/n))  \notag \\
&=&O(n^{\xi (x_{0})})  \label{un2}
\end{eqnarray}%
(cf.\ Corollary 1.2.4 and Remark 1.2.11 in \cite{deHaan07}). Then after
applying the triangle inequality and the smoothness\ and boundedness of $%
c(\cdot )$, $d(\cdot )$, and $\gamma (\cdot )$ (Condition B.(i)), we have
that for some constant $C>0,$%
\begin{eqnarray}
&&\sup_{x\in B_{\eta _{T}}(x_{0})}\left\vert \left\vert \frac{\partial
F_{Y|X=x}\left( u_{n}\right) /\partial x}{1-F_{Y|X=x}\left( u_{n}\right) }%
\right\vert \right\vert   \label{ineq1} \\
&\leq &\sup_{x\in B_{\eta _{T}}(x_{0})}\left\Vert \frac{c_{1}(x)}{c(x)}-\log
\left( u_{n}\right) \gamma _{1}(x)+Cd_{1}(x)(u_{n})^{-\tilde{\gamma}%
(x)}\right.   \notag \\
&&\left. -Cd(x)u_{n}^{-\tilde{\gamma}(x)}\tilde{\gamma}(x)\log
(u_{n})+Cr_{1}(x,u_{n})\right\Vert   \notag \\
&=&O(\log (u_{n}))\text{ (by Condition B.(i))}  \notag \\
&=&O(\log n).\text{ (by (\ref{un2}))}  \notag
\end{eqnarray}%
By (\ref{cond Bi1}) again, we have%
\begin{eqnarray}
&&\sup_{x\in B_{\eta _{T}}(x_{0})}\left\vert \frac{1-F_{Y|X=x}\left(
u_{n}\right) }{1-F_{Y|X=x_{0}}(u_{n})}\right\vert   \label{ineq2} \\
&\leq &\sup_{x\in B_{\eta _{T}}(x_{0})}\left\vert u_{n}^{-\gamma (x)+\gamma
(x_{0})}\right\vert \sup_{x\in B_{\eta _{T}}(x_{0})}\left\vert \frac{c(x)}{%
c(x_{0})}\right\vert \sup_{x\in B_{\eta _{T}}(x_{0})}\left\vert \frac{%
1+d(x)(y)^{-\tilde{\gamma}(x)}+r(x,u_{n})}{1+d(x_{0})(y)^{-\tilde{\gamma}%
(x_{0})}+r(x_{0},u_{n})}\right\vert   \notag \\
&\leq &C\exp \left( \sup_{x\in B_{\eta _{T}}(x_{0})}\log \left(
u_{n}^{-\gamma (x)+\gamma (x_{0})}\right) \right) \text{ \ \ (by Condition
B.(i))}  \notag \\
&=&C\exp \left( O(T^{-\eta }\log \left( u_{n}\right) )\right)   \notag \\
&=&C\exp (O(T^{-\eta }\log n)\text{ \ \ \ \ \ \ \ \ \ \ \ \ \ \ (by (\ref%
{un2}))}  \notag \\
&=&O(1).\text{ \ \ \ \ \ \ \ \ \ \ \ \ \ \ \ \ \ \ \ \ \ \ \ \ \ (by
Condition 1.4)}  \notag
\end{eqnarray}%
Then part (a) follows by combining (\ref{ineq1}) and (\ref{ineq2}) and using 
$O(T^{-\eta })\times O(\log n)=o(1)$ by Condition 1.4 again. }

{\small For (b), Condition B.(i) implies that 
\begin{eqnarray}
f_{Y|X=x}\left( y\right)  &=&-c(x)\gamma (x)(y)^{-\gamma (x)-1}(1+d(x)(y)^{-%
\tilde{\gamma}(x)}+r(x,y))  \label{fy} \\
&&+c(x)(y)^{-\gamma (x)}(-d(x)y^{-\tilde{\gamma}(x)-1}\tilde{\gamma}%
(x)+r_{2}(x,y)).  \notag
\end{eqnarray}%
A similar argument as above with Conditions B.(i) and 1.4 yields%
\begin{equation*}
\sup_{x\in B_{\eta _{T}}(x_{0})}\left\vert \left\vert \frac{\partial
f_{Y|X=x}\left( u_{n}\right) /\partial x}{f_{Y|X=x}\left( u_{n}\right) }%
\right\vert \right\vert \leq O(\log (u_{n}))=O(\log n)
\end{equation*}%
and 
\begin{equation*}
\sup_{x\in B_{\eta _{T}}(x_{0})}\left\vert \left\vert \frac{f_{Y|X=x}\left(
u_{n}\right) }{f_{Y|X=x_{0}}\left( u_{n}\right) }\right\vert \right\vert
\leq C\exp \left( \sup_{x\in B_{\eta _{T}}(x_{0})}\log \left( u_{n}^{-\gamma
(x)+\gamma (x_{0})}\right) \right) =O(1),
\end{equation*}%
which yield part (b) by using Condition 1.4 again. }

\bigskip {\small The proof for }$\xi (x_{0})<0${\small \ is very similar
when we replace }$y${\small \ with }$y_{0}-y${\small . In particular, by the
chain rule and the condition (Condition B.(iii)) that 
\begin{equation}
1-F_{Y|X=x}(y)=c(x)\left( y_{0}-y\right) ^{-\gamma (x)}(1+d(x)(y_{0}-y)^{-%
\tilde{\gamma}(x)}+r(x,y)),  \label{cond Bi2}
\end{equation}%
we have%
\begin{eqnarray*}
&&\frac{\partial F_{Y|X=x}\left( y\right) /\partial x}{1-F_{Y|X=x}\left(
y\right) } \\
&=&\frac{c_{1}(x)}{c(x)}-\gamma _{1}(x)\log \left( y_{0}-y\right) +\frac{%
d_{1}(x)\left( y_{0}-y\right) ^{-\tilde{\gamma}(x)}}{1+d(x)(y_{0}-y)^{-%
\tilde{\gamma}(x)}+r(x,y)} \\
&&-\frac{d(x)\left( y_{0}-y\right) ^{-\tilde{\gamma}(x)}\tilde{\gamma}%
_{1}(x)\log \left( y_{0}-y\right) }{1+d(x)(y_{0}-y)^{-\tilde{\gamma}%
(x)}+r(x,y)}+\frac{r_{1}(x,y)}{1+d(x)(y_{0}-y)^{-\tilde{\gamma}(x)}+r(x,y)}.
\end{eqnarray*}%
Similarly as (\ref{un2}), we denote }$u_{n}=a_{n}y+b_{n}$ and define $\bar{u}%
_{n}=y_{0}-u_{n}$. Then {\small \ 
\begin{eqnarray}
\bar{u}_{n} &=&y_{0}-u_{n}  \label{un3} \\
&=&\left( y_{0}-Q_{Y|X=x_{0}}(1\left. -\right. 1/n)\right) \left( 1+\xi
(x_{0})y\right)   \notag \\
&=&O(n^{\xi (x_{0})}),  \notag
\end{eqnarray}%
and then%
\begin{eqnarray}
&&\sup_{x\in B_{\eta _{T}}(x_{0})}\left\vert \left\vert \frac{\partial
F_{Y|X=x}\left( u_{n}\right) /\partial x}{1-F_{Y|X=x}\left( u_{n}\right) }%
\right\vert \right\vert   \label{ineq3} \\
&\leq &\sup_{x\in B_{\eta _{T}}(x_{0})}\left\Vert \frac{c_{1}(x)}{c(x)}-\log
\left( \bar{u}_{n}\right) \gamma _{1}(x)+Cd_{1}(x)(\bar{u}_{n})^{-\tilde{%
\gamma}(x)}\right.   \notag \\
&&\left. -Cd(x)\bar{u}_{n}^{-\tilde{\gamma}(x)}\tilde{\gamma}(x)\log (\bar{u}%
_{n})+Cr_{1}(x,u_{n})\right\Vert   \notag \\
&=&O(-\log (\bar{u}_{n}))\text{ (by Condition B.(iii))}  \notag \\
&=&O(\log n).\text{ (by (\ref{un3}))}  \notag
\end{eqnarray}%
}

{\small By (\ref{cond Bi2}) again, we have%
\begin{eqnarray}
&&\sup_{x\in B_{\eta _{T}}(x_{0})}\left\vert \frac{1-F_{Y|X=x}\left(
u_{n}\right) }{1-F_{Y|X=x_{0}}(u_{n})}\right\vert  \label{ineq4} \\
&\leq &\sup_{x\in B_{\eta _{T}}(x_{0})}\left\vert \bar{u}_{n}^{-\gamma
(x)+\gamma (x_{0})}\right\vert \sup_{x\in B_{\eta _{T}}(x_{0})}\left\vert 
\frac{c(x)}{c(x_{0})}\right\vert \sup_{x\in B_{\eta _{T}}(x_{0})}\left\vert 
\frac{1+d(x)(\bar{u}_{n})^{-\tilde{\gamma}(x)}+r(x,u_{n})}{1+d(x_{0})(\bar{u}%
_{n})^{-\tilde{\gamma}(x_{0})}+r(x_{0},u_{n})}\right\vert  \notag \\
&\leq &C\exp \left( \sup_{x\in B_{\eta _{T}}(x_{0})}\log \left( \bar{u}%
_{n}^{-\gamma (x)+\gamma (x_{0})}\right) \right) \text{ \ \ (by Condition
B.(iii))}  \notag \\
&=&C\exp \left( O(-T^{-\eta }\log \left( \bar{u}_{n}\right) )\right)  \notag
\\
&=&C\exp (O(T^{-\eta }\log n)\text{ \ \ \ \ \ \ \ \ \ \ \ \ \ \ (by (\ref%
{un3}))}  \notag \\
&=&O(1).\text{ \ \ \ \ \ \ \ \ \ \ \ \ \ \ \ \ \ \ \ \ \ \ \ \ \ (by
Condition 1.4)}  \notag
\end{eqnarray}%
Then part (a) follows by combining (\ref{ineq3}) and (\ref{ineq4}) and using 
$O(T^{-\eta })\times O(\log n)=o(1)$ by Condition 1.4 again. }

{\small For (b), Condition B.(iii) implies that 
\begin{eqnarray*}
f_{Y|X=x}\left( y\right)  &=&c(x)\gamma (x)(y_{0}-y)^{-\gamma
(x)-1}(1+d(x)(y_{0}-y)^{-\tilde{\gamma}(x)}+r(x,y)) \\
&&+c(x)(y_{0}-y)^{-\gamma (x)}(d(x)(y_{0}-y)^{-\tilde{\gamma}(x)-1}\tilde{%
\gamma}(x)+r_{2}(x,y)).
\end{eqnarray*}%
A similar argument as above with Conditions B.(iii) and 1.4 yields%
\begin{equation*}
\sup_{x\in B_{\eta _{T}}(x_{0})}\left\vert \left\vert \frac{\partial
f_{Y|X=x}\left( u_{n}\right) /\partial x}{f_{Y|X=x}\left( u_{n}\right) }%
\right\vert \right\vert \leq O(-\log (\bar{u}_{n}))=O(\log n)
\end{equation*}%
and 
\begin{equation*}
\sup_{x\in B_{\eta _{T}}(x_{0})}\left\vert \left\vert \frac{f_{Y|X=x}\left(
u_{n}\right) }{f_{Y|X=x_{0}}\left( u_{n}\right) }\right\vert \right\vert
\leq C\exp \left( \sup_{x\in B_{\eta _{T}}(x_{0})}\left( -\log \left( \bar{u}%
_{n}^{-\gamma (x)+\gamma (x_{0})}\right) \right) \right) =O(1),
\end{equation*}%
which yield part (b) by using Condition 1.4 again. }

{\small Now it remains prove (a) and (b) for $\xi (x_{0})\left. =\right. 0$.
Note that $u_{n}=O\left( Q_{Y|X=x_{0}}(1-1/n)\right) $, which is at most of
the order $\exp (\Phi ^{-1}(1-1/n))=\exp (\sqrt{2\log n})$ by the condition $%
C_{1}(\log y)^{2}\leq \tilde{d}(y)\leq C_{2}y^{C_{3}}$. }

{\small For (a), we decompose $B_{\eta _{T}}(x_{0})$ into $B_{\eta
_{T}}(x_{0})\left. \cap \right. \{x\left. :\right. \xi (x)\left. >\right.
0\} $, $B_{\eta _{T}}(x_{0})\left. \cap \right. \{x\left. :\right. \xi
(x)\left. =\right. 0\}$, and $B_{\eta _{T}}(x_{0})\left. \cap \right.
\{x\left. :\right. \xi (x)\left. <\right. 0\}$ and then 
\begin{footnotesize}
\begin{eqnarray}
&&\sup_{x\in B_{\eta _{T}}(x_{0})}\left\vert \left\vert \frac{\partial
F_{Y|X=x}\left( u_{n}\right) /\partial x}{1-F_{Y|X=x_{0}}\left( u_{n}\right) 
}\right\vert \right\vert \leq \max \left\{ \sup_{x\in B_{\eta
_{T}}(x_{0})\cap \{x:\xi (x)>0\}}\left\vert \left\vert \frac{\partial
F_{Y|X=x}\left( u_{n}\right) /\partial x}{1-F_{Y|X=x_{0}}\left( u_{n}\right) 
}\right\vert \right\vert ,\right.  \notag \\
&&\left. \sup_{x\in B_{\eta _{T}}(x_{0})\cap \{x:\xi (x)=0\}}\left\vert
\left\vert \frac{\partial F_{Y|X=x}\left( u_{n}\right) /\partial x}{%
1-F_{Y|X=x_{0}}\left( u_{n}\right) }\right\vert \right\vert ,\sup_{x\in
B_{\eta _{T}}(x_{0})\cap \{x:\xi (x)<0\}}\left\vert \left\vert \frac{%
\partial F_{Y|X=x}\left( u_{n}\right) /\partial x}{1-F_{Y|X=x_{0}}\left(
u_{n}\right) }\right\vert \right\vert .\right\}  \label{max decom}
\end{eqnarray}%
\end{footnotesize}%
For the first item in (\ref{max decom}), Conditions 1.1 and B.(i) imply that 
$\partial F_{Y|X=x}(u_{n})/\partial x=O(u_{n}^{-\gamma (x)}\log u_{n})$ and $%
\gamma (x)=1/\xi (x)=1/\xi ^{\prime }(\dot{x})T^{\eta }\geq O(T^{\eta })$
where $\dot{x}$ is within $B_{\eta _{T}}(x_{0})$ such that }${\small 1/\xi }%
^{\prime }{\small (\dot{x})>0}${\small . Thus, Condition 1.4 and the fact
that $1-F_{Y|X=x_{0}}\left( u_{n}\right) =O(n^{-1})$ yield that for any $%
x\in B_{\eta _{T}}(x_{0})\cap \{x:\xi (x)>0\}$, 
\begin{eqnarray*}
&&\left\vert \left\vert \frac{\partial F_{Y|X=x}\left( u_{n}\right)
/\partial x}{1-F_{Y|X=x_{0}}\left( u_{n}\right) }\right\vert \right\vert \\
&=&O\left( n\times u_{n}^{-\gamma (x)}\log u_{n}\right) \\
&=&O\left( \exp \left( \log n-\gamma (x)\log u_{n}+\log \left( \log
u_{n}\right) \right) \right) \\
&\leq &O(\exp \left( \log n-T^{\eta }\log u_{n}+\log \left( \log
u_{n}\right) \right) ) \\
&=&o(1).
\end{eqnarray*}%
}

{\small For the second term in (\ref{max decom}), apply Leibniz's rule and
Condition B.(ii) to obtain 
\begin{eqnarray}
&&\sup_{x\in B_{\eta _{T}}(x_{0})\cap \{x:\xi (x)=0\}}\left\vert \left\vert 
\frac{\partial F_{Y|X=x}\left( u_{n}\right) /\partial x}{1-F_{Y|X=x_{0}}%
\left( u_{n}\right) }\right\vert \right\vert  \notag \\
&\leq &\sup_{x\in B_{\eta _{T}}(x_{0})\cap \{x:\xi
(x)=0\}}Cn\int_{u_{n}}^{y_{0}}y^{C_{3}}f_{Y|X=x}(y)dy  \notag \\
&\leq &Cn\int_{u_{n}}^{y_{0}}y^{C_{3}+\bar{C}_{T}}\exp (-\underline{D}%
_{T}\left( \log y\right) ^{2})dy  \label{bound tail} \\
&=&Cn\int_{\log u_{n}}^{y_{0}}\exp (-\underline{D}_{T}s^{2}+(C_{3}+\bar{C}%
_{T}+1)s)ds\text{ (by change of variables)}  \notag \\
&=&O(1),  \notag
\end{eqnarray}%
where we denote $\bar{C}_{T}\left. =\right. \sup_{x\in B_{\eta _{T}}(x_{0})}%
\tilde{c}(x)\left. <\right. \infty $ and $\underline{D}_{T}\left. =\right.
\inf_{x\in B_{\eta _{T}}(x_{0})}d(x)\left. >\right. 0$, and the last
equation follows from that $u_{n}$ is at most of the order $\exp (\sqrt{%
2\log n})$ and the fact that the $1\left. -\right. 1/n$ quantile of a normal
distribution is $O(\sqrt{\log (n)})$. }

For the third term in (\ref{max decom}), Conditions 1.1 and B.(iii) imply%
{\small \ that $\partial F_{Y|X=x}(u_{n})/\partial x=O(-\bar{u}_{n}^{-\gamma
(x)}\log \bar{u}_{n})$ and }$-{\small \gamma (x)}\left. {\small =}\right. $%
{\small $-1/\xi (x)=-1/\xi ^{\prime }(\dot{x})T^{\eta }\geq O(T^{\eta })$
where $\dot{x}$ is within $B_{\eta _{T}}(x_{0})$ such that }${\small -1/\xi }%
^{\prime }{\small (\dot{x})>0}$. {\small Thus, Condition 1.4 and the fact
that $1-F_{Y|X=x_{0}}\left( u_{n}\right) =O(n^{-1})$ yield that for any $%
x\in B_{\eta _{T}}(x_{0})\cap \{x:\xi (x)<0\}$, 
\begin{eqnarray*}
&&\left\vert \left\vert \frac{\partial F_{Y|X=x}\left( u_{n}\right)
/\partial x}{1-F_{Y|X=x_{0}}\left( u_{n}\right) }\right\vert \right\vert \\
&=&O\left( n\times \bar{u}_{n}^{-\gamma (x)}\left( -\log \bar{u}_{n}\right)
\right) \\
&=&O\left( \exp \left( \log n-\gamma (x)\log \bar{u}_{n}+\log \left( -\log 
\bar{u}_{n}\right) \right) \right) \\
&\leq &O(\exp \left( \log n + T^{\eta }\log \bar{u}_{n}+\log \left( -\log \bar{%
u}_{n}\right) \right) )\text{ (by }\log \bar{u}_{n}<0\text{)} \\
&=&o(1).
\end{eqnarray*}%
}

{\small For (b), we similarly derive%
\begin{footnotesize}
\begin{eqnarray}
&&\sup_{x\in B_{\eta _{T}}(x_{0})}\left\vert \left\vert \frac{\partial
f_{Y|X=x}\left( u_{n}\right) /\partial x}{f_{Y|X=x_{0}}\left( u_{n}\right) }%
\right\vert \right\vert \leq \max \left\{ \sup_{x\in B_{\delta
_{T}}(x_{0})\cap \{x:\xi (x)>0\}}\left\vert \left\vert \frac{\partial
f_{Y|X=x}\left( u_{n}\right) /\partial x}{f_{Y|X=x_{0}}\left( u_{n}\right) }%
\right\vert \right\vert ,\right.   \label{max f decom} \\
&&\left. \sup_{x\in B_{\delta _{T}}(x_{0})\cap \{x:\xi (x)=0\}}\left\vert
\left\vert \frac{\partial f_{Y|X=x}\left( u_{n}\right) /\partial x}{%
f_{Y|X=x_{0}}\left( u_{n}\right) }\right\vert \right\vert ,\sup_{x\in
B_{\delta _{T}}(x_{0})\cap \{x:\xi (x)<0\}}\left\vert \left\vert \frac{%
\partial f_{Y|X=x}\left( u_{n}\right) /\partial x}{f_{Y|X=x_{0}}\left(
u_{n}\right) }\right\vert \right\vert \right\} .  \notag
\end{eqnarray}%
\end{footnotesize}%
Using (\ref{fy}) and Condition B.(i), we have $\left\vert \left\vert
\partial f_{Y|X=x}\left( u_{n}\right) /\partial x\right\vert \right\vert
=O\left( u_{n}^{-\gamma \left( x\right) -1}\gamma \left( x\right) \right)
\left. +\right. O\left( u_{n}^{-\gamma \left( x\right) }\right) $ when $\xi
(x)\left. >\right. 0$. By Condition B.(ii) and under $\xi (x_{0})\left.
=\right. 0$, we have $1/f_{Y|X=x_{0}}(u_{n})\left. \leq \right. Cu_{n}\exp
\left( \bar{D}_{T}C_{2}u_{n}^{C_{3}}\right) $ where we denote $\bar{D}%
_{T}\left. =\right. \sup_{x\in B_{\eta _{T}}\left( x_{0}\right) }d(x)\left.
>\right. 0$. Thus for any $x\left. \in \right. B_{\delta _{T}}(x_{0})\left.
\cap \right. \{x\left. :\right. \xi (x)\left. >\right. 0\}$,%
\begin{eqnarray*}
&&\left\vert \left\vert \frac{\partial f_{Y|X=x}\left( u_{n}\right)
/\partial x}{f_{Y|X=x_{0}}\left( u_{n}\right) }\right\vert \right\vert  \\
&\leq &Cu_{n}\exp \left( \bar{D}_{T}C_{2}u_{n}^{C_{3}}\right) \left(
u_{n}^{-\gamma \left( x\right) -1}\gamma \left( x\right) +u_{n}^{-\gamma
\left( x\right) }\right)  \\
&=&Cu_{n}\exp \left( \bar{D}_{T}C_{2}u_{n}^{C_{3}}-(\gamma \left( x\right)
+1)\log \left( u_{n}\right) +\log \gamma \left( x\right) \right)  \\
&&+u_{n}C\exp \left( \bar{D}_{T}C_{2}u_{n}^{C_{3}}-\gamma \left( x\right)
\log \left( u_{n}\right) \right)  \\
&\leq &Cu_{n}\exp \left( \bar{D}_{T}C_{2}u_{n}^{C_{3}}-T^{-\eta }\log \left(
u_{n}\right) +\log \gamma \left( x\right) \right)  \\
&=&o(1),
\end{eqnarray*}%
where the last line follows from Condition 1.4 and the fact that $u_{n}$ is
at most of the order $\exp (\sqrt{2\log n}).$ }

{\small The second term in (\ref{max f decom}) is bounded by 
\begin{eqnarray*}
&&\sup_{x\in B_{\eta _{T}}(x_{0})}\left\vert \left\vert \frac{c_{1}(x)}{c(x)}%
+\frac{1}{u_{n}}\tilde{c}_{1}(x)+d_{1}(x)\tilde{d}(u_{n})+\frac{%
r_{1}(x,u_{n})}{1+r(x,u_{n})}\right\vert \right\vert  \\
&\leq &O(u_{n}^{C_{3}})\leq O(\left( \log (n)\right) ^{C_{3}/2}).
\end{eqnarray*}%
}

{\small To bound the the third term in (\ref{max f decom}), we have }$%
\left\vert \left\vert \partial f_{Y|X=x}\left( u_{n}\right) /\partial
x\right\vert \right\vert =O\left( -\bar{u}_{n}^{-\gamma \left( x\right)
-1}\gamma \left( x\right) \right) \left. +\right. O\left( \bar{u}%
_{n}^{-\gamma \left( x\right) }\right) ${\small \ when }$\xi (x)\left.
<\right. 0${\small . Then similarly as bounding the first term, we have%
\begin{eqnarray*}
&&\left\vert \left\vert \frac{\partial f_{Y|X=x}\left( u_{n}\right)
/\partial x}{f_{Y|X=x_{0}}\left( u_{n}\right) }\right\vert \right\vert  \\
&\leq &Cu_{n}\exp \left( \bar{D}_{T}C_{2}u_{n}^{C_{3}}\right) \left( -\bar{u}%
_{n}^{-\gamma \left( x\right) -1}\gamma \left( x\right) +\bar{u}%
_{n}^{-\gamma \left( x\right) }\right)  \\
&=&Cu_{n}\exp \left( \bar{D}_{T}C_{2}u_{n}^{C_{3}}-(\gamma \left( x\right)
+1)\log \left( \bar{u}_{n}\right) +\log \left( -\gamma \left( x\right)
\right) \right)  \\
&&+u_{n}C\exp \left( \bar{D}_{T}C_{2}u_{n}^{C_{3}}-\gamma \left( x\right)
\log \left( \bar{u}_{n}\right) \right)  \\
&\leq &Cu_{n}\exp \left( \bar{D}_{T}C_{2}u_{n}^{C_{3}}+T^{-\eta }\log \left(
\bar{u}_{n}\right) +\log \left( -\gamma \left( x\right) \right) \right)  \\
&=&o(1),
\end{eqnarray*}%
Thus (b) for $\xi (x_{0})\left. =\right. 0$ is established. $\blacksquare $ }

\renewcommand{\baselinestretch}{1.3} {\small \vspace{.1cm} }

{\normalsize 
\bibliographystyle{econometrica}
\bibliography{diss}
}

\end{document}